\documentclass[%
 reprint,
superscriptaddress,
 amsmath,amssymb,
 aps,
floatfix
]{revtex4-2}

\usepackage{graphicx}
\usepackage{float}
\usepackage{dcolumn}
\usepackage{bm}
\usepackage{hyperref}
\usepackage[mathlines]{lineno}
\usepackage{comment}
\usepackage{caption}
\usepackage{subcaption}

\usepackage{xcolor}

\usepackage[misc]{ifsym}

\begin{document}

\preprint{APS/123-QED}

\title{The effect of a linear feedback mechanism in a homeostasis model}

\author{Antonio Francesco Zirattu}\email{Corresponding author: antoniofrancesco.zirattu@unito.it}
\author{Marta Biondo}
\author{Matteo Osella}
\author{Michele Caselle}

\affiliation{
University of Turin - Physics Department and INFN - Section of Turin\\ Via Pietro Giuria 1, 10125, Turin
}


\date{\today}


\begin{abstract}
\noindent Feedback loops are essential for regulating cell proliferation and maintaining the delicate balance between cell division and cell death. 
Thanks to the exact solution of a few simple models of cell growth it is by now clear that stochastic fluctuations play a central role in this process and that cell growth (and in particular the robustness and stability of homeostasis) can be properly addressed only as a stochastic process.
Using epidermal homeostasis as a prototypical example, we show that it is possible to discriminate among different feedback strategies which turn out to be characterized by  different, experimentally testable, behaviours.
In particular, we focus on the so-called Dynamical Heterogeneity model, an epidermal homeostasis model that takes into account two well known cellular features: the plasticity of the cells and their adaptability to face environmental stimuli. We show that specific choices of the parameter on which the feedback is applied may decrease the fluctuations of the homeostatic population level and improve the recovery of the system after an external perturbation. 
\end{abstract}

\maketitle
\section{\label{sec:level1} Introduction}
\footnotetext{
* E-mail: \href{mailto:antoniofrancesco.zirattu@unito.it}{antoniofrancesco.zirattu@unito.it}}
\noindent  Cell proliferation is a critical aspect of growth, development, and tissue repair. It is also a complex and finely regulated process that must be maintained within specific boundaries to avoid uncontrolled cell growth and the formation of tumors. To ensure proper regulation of cell proliferation, feedback loops play a crucial role in maintaining the delicate balance between cell division and cell death. Feedback control can be implemented in different ways, depending on the model of interest and on the particular process on which the feedback control acts. \\
The main goal of this paper is to show that these different choices are not on the same ground: while all of them ensure homeostasis, they differ in the rapidity with which homeostatic levels are recovered after a perturbation and in the effectiveness in controlling the variability of steady state population and of the clone survival rate. This makes it possible, at least in principle, to identify from experimental data which mechanism is at work in a particular cell model and, more importantly, to guess which robustness requirement led the system to select a particular feedback mechanism among the others.\\
Thanks to the exact solution of a few simple homeostasis models \cite{antal2010exact} it is by now clear that stochastic fluctuations play a central role in  this game and that models which are apparently stable on a deterministic level cannot grant a robust homeostasis if stochastic fluctuations are taken into account. It is thus mandatory to perform this analysis using stochastic equations and, as we shall see, stability with respect to stochastic fluctuations will be a major constraint in selecting viable models of homeostasis.\\
In the following we shall concentrate in particular on epidermal homeostasis, for which several experimental models exist, but our results are of general nature and could be applied with suitable changes to any model describing stable homeostatic states in growing cell populations. 
\subsection{\label{Epidermal_homeostasis}Epidermal homeostasis}

\noindent The epidermis is the outermost layer of the skin and it is responsible for key protective, secretory and regulatory functions \cite{mcgrath2004anatomy}. From the histological point of view, the epidermis is a stratified epithelium mostly populated by keratinocytes, cells that are continuously produced in the basal layer, the \textit{stratum basale}, and shed from the outermost layer, termed \textit{stratum corneum}. In order to preserve the structure and the functionality of the tissue, \textit{i.e.} its homeostasis, the balance between the numbers of lost and newborn cells must be finely regulated. Understanding all the mechanisms that contribute to the maintenance of the skin homeostasis is crucial because its dysregulation leads to severe health problems such as skin cancer \cite{augustin2015wnt}\cite{garg2001psychological}.\newline
It is well known that regenerative tissues such as the epidermis present a specific class of cells termed \textit{stem cells}, which are able to self-renew and to generate various lineages of differentiated cells \cite{potten1996stem}. The role of stem cells in the epidermis has been debated for a long time and has not been fully clarified yet. \newline
The first mathematical model of epidermal homeostasis assumed that the epidermis is composed of distinct proliferative units, each maintained by a single stem cell that renews itself and generates a transit-amplifying (TA) cell upon cellular division \cite{allen1974fine}\cite{potten1981cell}. In turn, TA cells undergo cell division a few times before producing fully differentiated keratinocytes that eventually migrate towards the upper layers of the tissue. In this model, the homeostasis is maintained at the single-level through asymmetric stem cell divisions, which guarantee the preservation of a constant number of stem cells.\newline
After the establishment of lineage tracing methods \cite{kretzschmar2012lineage}, new experimental data prompted the proposal of a different paradigm of epidermal homeostasis, described by the so-called hierarchical model (H model). In this model, the homeostasis is maintained at the whole cellular population level by an entirely stochastic hierarchical process \cite{clayton2007single}\cite{dekoninck2020defining} that involves two populations of cells: progenitor cells ($A$), able to divide and proliferate, and differentiated cells ($B$), bound to leave the basal layer and populate the tissue. The outcome of the cellular divisions is stochastic and may be either symmetric ($A \ \xrightarrow{}  A  +  A$ or $A  \xrightarrow{} B  +  B$) or asymmetric ($A \xrightarrow{} A  +  B$). \newline 
The discrepancy between these two models was overcome in 2012 thanks to new experimental data, which prompted the development of a third model that brings together the presence of a slow cycling population of stem cells, that renew and support a larger pool of proliferative cells that divide and produce differentiated cells \cite{mascre2012distinct}\cite{lei2020general}.

\subsection{\label{Cellular_plasticity_and_robustness_issues}Cellular plasticity and robustness issues of the epidermal homeostasis models}

\noindent Despite being inherently different, the three aforementioned models share two features: firstly, they all suffer from a constitutive instability with respect to any possible fluctuations of their parameters. Secondly, the cells are organized in a one-way hierarchical structure, which does not leave room for a relevant, well known cellular characteristic termed \textit{cell plasticity} \cite{blanpain2014plasticity}.\newline
Cell plasticity is the capability of cells primed to differentiation to reacquire proliferative potential both in normal \cite{doupe2012single, ritsma2014intestinal, hara2014mouse, kai2004differentiating} and pathological \cite{donati2015stem, alcolea2014differentiation, morgan2018targeting, shimokawa2017visualization} conditions, as well as in wound repairing processes \cite{tata2013dedifferentiation, roshan2016human}.\newline
Regarding the lack of robustness, the homeostasis must be robust to the possible failure of some of the underlying microscopic processes that preserve it. The existence of control mechanisms in growing and developed tissues is well known, and it is paramount to achieve and preserve the desired shape and size of adult organs and tissues \cite{lander2009cell}. Hence, the stability of the system with respect to stochastic fluctuations of the parameters should be taken into account in all epidermal homeostasis models.\newline
The three aforementioned models correspond to three critical, structurally unstable branching processes \cite{asmussen1983branching, strogatz2018nonlinear}, which can be rendered robust through the introduction of a suitable feedback mechanism. As a matter of fact, it is well known that body cells can adjust their behaviour to face changes in the surrounding environment or in response to mechanical, chemical or mechanochemical signals \cite{shraiman2005mechanical, alric2022macromolecular}. \newline
In particular, epidermal cells adjust their division and  loss rates if the surrounding environment becomes overcrowded: this mechanism is based on contact inhibition \cite{puliafito2012collective} and is termed \textit{crowding feedback} \cite{eisenhoffer2013bringing, marinari2012live}. Despite many efforts and a considerable amount of acquired knowledge on the topic, there is a lot more to unveil concerning the exact molecular pathways that contribute to the feedback mechanisms and how they quantitatively affect the behaviour of the cells \cite{schiefelbein2014regulation, rauch2016mapk, chan2017coordination}.\newline
Both cell plasticity and the robustness issues have been addressed by Greulich et al. \cite{greulich2016dynamic}, who proposed a modified version of the H model that takes into account the \textit{dynamical heterogeneity} (DH) of the cells, \textit{i.e.} the capability to revert their state from differentiated to proliferative. This model, termed DH model, is fully described by four parameters that correspond to the cell division, death and change of state (from proliferative to differentiated and vice versa) rates (see Material and Methods).\newline
A deterministic analysis of the model suggests that any choice of the parameter subject to feedback would be equally good in conferring robustness to the model. Instead, we question how each specific choice affects relevant aspects of the dynamics of the system, which is inherently stochastic: the survival probability of the cell population in a growth process, the variability of the cellular population at the stationary, homeostatic state, and the recovery time after an injury event.\newline
Concerning the first two aspects, the feedback must enhance the survival probability of clones that arise from a single or a few cells, in which a temporary imbalance of the cellular parameters towards proliferation allow to reach the desired number of cells\cite{dekoninck2020defining}. Afterwards, once the stationary state is reached, the presence of control mechanisms must make the variability of the cellular number around it to be low \cite{kok2022minimizing, huelsz2022mother}. \newline
Finally, speaking of the recovery time, it is the time it takes the cellular population to recover from an injury event. Mathematically speaking, the recovery time is a \textit{first passage time} (FPT), which is a random variable equal to the time it takes a certain stochastic quantity (in this case the number of cells in a clone) to reach a specific value \cite{gardiner1985handbook}. FPTs are crucial quantities for the biology of the cell because of the inherent randomness of most of the processes that take place inside it, such as the diffusion of molecules that have to reach their target in time \cite{stein2012transport}, or the concentration level of proteins which must be high enough to initiate specific processes such as protein transcription \cite{co2017stochastic, alon2006introduction}. A correct timing is crucial in healing processes, hence it would be noticeable if a specific feedback mechanism stabilized or sped up the recovery time of the system. \newline
In spite of being conducted in the context of the epidermis, our analysis could be extended to all the biological and non-biological systems whose behaviour is stochastic and in which the presence of some sort of control mechanism is indispensable.

\section{Material and methods}\label{Matherial_Methods}

\subsection*{The Hierarchical model}\label{SEC:3_1}
\noindent The hierarchical (H) model was introduced in 2007 by Clayton et al. \cite{clayton2007single} and was solved analytically a few years later by Antal and Krapivsky \cite{antal2010exact}. \\
This model deals with how homeostasis is preserved in murine tail epidermis, showing that experimental data are coherent with a simple model in which a single population of progenitor cells is able to maintain the interfollicular epidermis (IFE). The experiments were conducted by marking single cells with fluorescent proteins, by mean of the \textit{lineage tracing} technique \cite{kretzschmar2012lineage}, which allowed to follow the fate of a \textit{clone}, that is a labeled cell and its progeny.\\ The key feature of the H model is that the fate of the cells is entirely stochastic, which makes all of them, rather than a restricted pool of stem cells, collectively responsible for maintaining the homeostasis.\\
The experiments showed that most of the marked clones disappeared after a few weeks and that the size of the surviving ones increased linearly with time. To reproduce this result, the authors proposed a hierarchical model in which the homeostasis is maintained by a single population of progenitor cells that can divide either symmetrically to undergo self-renewal or asymmetrically to give birth to post-mitotic cells that will leave the basal layer. The H model is summarized by the following rate equations:

\begin{equation}\label{eq:1}
\begin{split}
    A \  &\xrightarrow[]{\lambda_H} \left\{
    \begin{array}{l}
    A \ + \ A \quad\quad \left(\text{with probability} \ r \right)
    \\
    A \ + \ B \quad\quad \left(\text{with probability} \ 1  -  2r \right)
    \\
    B \ + \ B \quad\quad \left(\text{with probability} \ r  \right)
    \end{array}\right.
    \\
    B \ &\xrightarrow[]{\gamma_H} \ \varnothing
\end{split}
\end{equation}

\noindent $A$ represents a proliferative cell, $B$ stands for a differentiated cell, $\lambda_H$ is the division rate of proliferative cells and $\gamma_H$ is the loss rate of differentiated cells. $r$ is the probability that the division of a proliferative cell is symmetric, \textit{i.e.} the division produces two equal cells. Conversely, $(1  -  2r)$ is the probability of an asymmetric cell division. \\ 
To preserve a constant population, the number of cell divisions must equalize the number of lost differentiated cells, \textit{i.e.} $\rho_H\lambda_H  =  \gamma_H(1-\rho_H)$, where $\rho_H$ is the proportion of progenitor cells in the basal layer. \\
The fit of experimental data provided the following values for the parameters of the model \cite{clayton2007single}:

\begin{itemize}
    \item $\lambda_H = 1.1\ week^{-1}$;
    \item $\gamma_H = 0.31\ week^{-1}$;
    \item $r = 0.08$;
\end{itemize}

\noindent Let $P_{N_A,N_B}(t)$ be the probability that a clone contains $N_A$ type $A$ cells and $N_B$ type $B$ cells at time $t$; $P_N(t)$ be the probability that a clone contains $N = N_A+N_B$ cells at time $t$. It can be shown that the long term behaviour of the size distributions depends on $\rho_H$ and $r\lambda_H$. For an initial condition:
\begin{equation*}
\begin{split}
    P_{N_A,N_B}(t=0) = \ &\rho_H\delta_{(N_A,1)}\delta_{(N_B,0)} + 
    \\
    +&(1-\rho_H)\delta_{(N_A,0)}\delta_{(N_B,1)}
\end{split}
\end{equation*}

\noindent(where $\delta_{(\cdot,\cdot)}$ is the Kronecker delta), the extinction probability and the asymptotic distribution of the total population of a clone are:

\begin{equation}\label{eq:2}
    \begin{split}
    \left\{
    \begin{array}{l}
    P_0(t) = \left(1 + \dfrac{1}{\Omega_H t}  \right)^{-1} =  \dfrac{\Omega_H t}{1 \ +\ \Omega_H t}
    \\
    \\
    \lim_{t\gg r\lambda_H} P_N^{surv}(t) = \dfrac{1}{\Omega_H t}\exp{-\dfrac{N}{\Omega_H t}}
    \end{array}\right.
    \end{split}
\end{equation}

\noindent Where $\Omega_H = \dfrac{r\lambda_H}{\rho_H} \ = \ 0.4 \  week^{-1}$.

\begin{figure}[h]
    \centering
    \includegraphics[width=0.95\linewidth, width=0.95\linewidth]{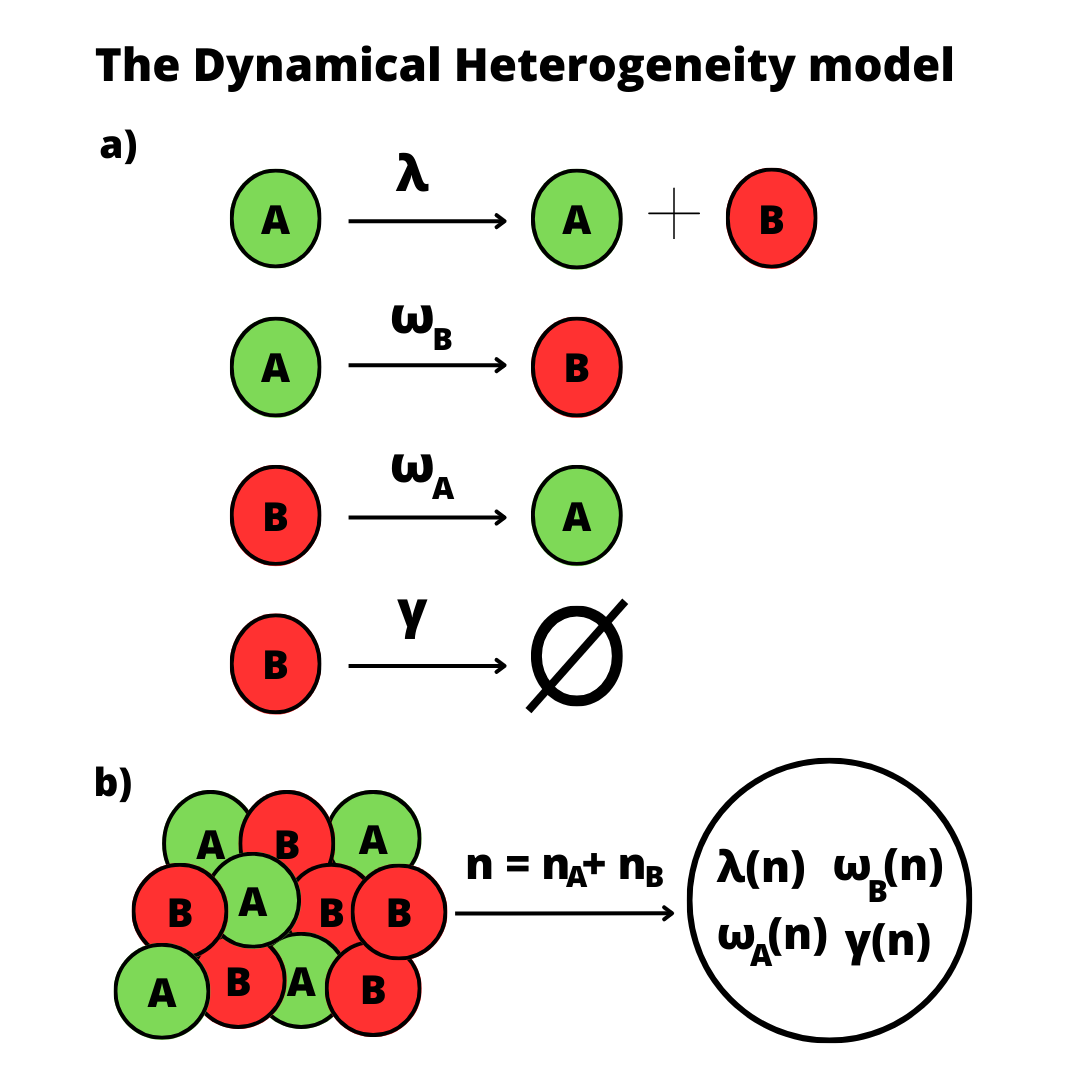}
    \caption{a) A schematic representation of the Dynamical Heterogeneity model. A reaction rate is associated to each possible reaction occurring in the model. \newline
    b) The total cellular number (or density) $n$ affects the reaction rates of the cells. This mechanism is termed crowding feedback and makes the homeostasis robust with respect to stochastic fluctuations of both the total cell number and the reaction rates (which depend on microscopic cellular processes).}
    \label{FIG_0}
\end{figure}

\subsection*{The Dynamical Heterogeneity model}\label{SEC:3_2}
\noindent The evidence of differentiated cells that may revert their state to a proliferative one, \textit{i.e.} cellular plasticity, prompted the development of another epidermal homeostasis model, the dynamical heterogeneity (DH) model \cite{greulich2016dynamic}, that extends the H model to take this feature into account.\\
The authors showed that this state-reversal cellular process is able to guarantee homeostasis on the macroscopic level alike a one-way hierarchy of cells does. Moreover, the authors discussed the robustness of the H and of the DH models to possible fluctuations of the parameters that regulate homeostasis: they showed that the implementation of a crowding feedback mechanism enhanced the stability of such models.\newline
The DH model is summarized by the following reactions:

\begin{equation}\label{eq:3}
\begin{split}
    A \  &\xrightarrow[]{\lambda} \ A\ +\ B
    \\
    A \  &\xrightarrow[]{\omega_B}  B
    \\
    B \  &\xrightarrow[]{\omega_A}  A
    \\
    B \ &\xrightarrow[]{\gamma} \ \varnothing
\end{split}
\end{equation}

\noindent Where:

\begin{itemize}
    \item $A$ stands for a progenitor cell, \textit{i.e.} one that can undergo mitosis and generate daughter cells. $B$ represents cell primed to differentiation and migration toward suprabasal layers;
    \item $\lambda$ is the division rate of type $A$ cells;
    \item $\gamma$ is the loss rate of type $B$ cells;
    \item $\omega_A$($\omega_B$) is the switching rate from state $B$($A$) to $A$($B$).
\end{itemize}

\noindent The time evolution of the average number of cells in a clone is given by the following system of differential equations, which can be derived by the master equation of the model:

\begin{equation}\label{eq:4}
\begin{split}
    \left\{
    \begin{array}{l}
    \dfrac{\partial N_A}{\partial t} = -\left( \omega_B + \omega_A\right)N_A  +  \omega_AN  
    \\
    \\
    \ \dfrac{\partial N}{\partial t}  = \left( \lambda  +  \gamma\right)N_A  -  \gamma N
    \end{array}\right.
\end{split}
\end{equation}

\noindent Where $N_A$ is the average number of type A cells and $N$ is the average total number of cells per clone. It can be readily checked that the system admits an homeostatic stationary state only if the parameters satisfy the following balance condition:

\begin{equation}\label{eq:5}
    \dfrac{\lambda}{\gamma}  -  \dfrac{\omega_B}{\omega_A}  =  0
\end{equation}

\noindent If Eq. (\ref{eq:5}) is verified, then $N_A(t)  =  N_{A}(t=0)  =  N_{A,0}$ and $N(t)  =  N(t=0)  =  N_{0}$. Moreover, it is possible to show that the H model and the DH model generate the same asymptotic distribution of the average size of surviving clones, reported in eq. (\ref{eq:2}). The only difference is the specific form of the scaling parameter $\Omega_{DH}$: 
\begin{equation*}
    \Omega_{DH} \ = \ \dfrac{\omega_B}{\rho}\left(1+\dfrac{\omega_B}{\lambda}\right) \left(1+\dfrac{\omega_B}{\rho\lambda}\right)^{-2}
\end{equation*}
\noindent Where $ \rho \ = \ \dfrac{\omega_A}{\omega_A+\omega_B} \ = \ \left(1+\dfrac{\lambda}{\gamma}\right)^{-1}$.

\subsection*{Crowding feedback}\label{SEC:3_3}
\noindent The reaction rates of the DH model must satisfy the balance condition reported in eq.(\ref{eq:5}) to reproduce a homeostatic state. If eq.(\ref{eq:5}) does not hold, then the average population size either diverges or goes extinct. However, the reactions rates are in general stochastic and vary from cell to cell, therefore we cannot expect the balance condition to be strictly satisfied. Nevertheless, a homeostatic system must be robust to stochastic events and respond to them accordingly. In fact, it is well known that the cells can adjust their response to a number of signals of diverse nature, and it has been observed that the division rate diminishes in response to an overcrowded environment: this phenomenon, termed contact inhibition, is an example of crowding feedback. \newline
In mathematical terms, a generic parameter $\theta$ is subject to crowding feedback if it depends on the average total cell number $N$: $\theta = \theta(N)$. In particular, $\dfrac{d\theta(N)}{dN}>0$ means that the overcrowding fosters the reactions whose $\theta$ is the rate of, whereas $\dfrac{d\theta(N)}{dN}<0$ means that the overcrowding hinders those reactions.\newline
To introduce the feedback in the DH model, it can be assumed that there is only one value of the average cell number such that the parameter subject to the crowding feedback satisfies the balance condition (keeping the other parameters fixed). This value will be denoted as $N^*$. To find the condition under which the crowding feedback mechanism makes the system robust, the linear stability analysis of the system can be performed. \newline
Let $\lambda$ be the parameter subjected to the feedback. Under this hypothesis, it can be shown that the system has a single stationary state $\Vec{X^*} = \left(\begin{smallmatrix} N_A^* \\ N^* \end{smallmatrix}\right) = \left(\begin{smallmatrix} \rho N^* \\ N^* \end{smallmatrix}\right)$.
Moreover, $\Vec{X^*}$ is linearly stable if and only if $\dfrac{d\lambda(N)}{dN} \ < \ 0$. If any of the other three parameters is chosen as subjected to the feedback, one would find out that:
\begin{enumerate}
    \item If $\omega_A = \omega_A(N)$, $\Vec{X^*}$ is stable if $\dfrac{d\omega_A(N)}{dN}<0$.
    \item If $\omega_B = \omega_B(N)$, $\Vec{X^*}$ is stable if $\dfrac{d\omega_B(N)}{dN}>0$.
    \item If $\gamma = \gamma(N)$, $\Vec{X^*}$ is stable if $\dfrac{d\gamma(N)}{dN}>0$.
\end{enumerate}

\noindent These requirements can be understood by looking at the physical meaning of the parameters: $\lambda$ and $\omega_A$ are the rates of the two reactions that either preserve or increase the number of progenitor cells in the system. Hence, if the system gets overcrowded or depleted their rate has to temporarily decrease or increase to regain the homeostatic state, respectively. Conversely, $\gamma$ and $\omega_B$ are the rates of the two reactions which either decrease the number of proliferative cells or the overall cell population, therefore their dependence on the total number of cells must vary in the opposite manner.

\subsection*{Simulations of the Dynamical Heterogeneity model}\label{SEC:3_4}

\subsubsection*{The choice of the parameters}\label{SEC:3_4_1}
\noindent Currently, there are no experimental data to fit the value of the four parameters of the DH model. Nevertheless, we exploited the similarity between the DH model and the H model and the fact that the latter's parameters were obtained through the fit of experimental data to set the former's ones.\newline This procedure was done as follows:
\begin{itemize}
    \item We chose to preserve the value of the death rate across the two models, i.e. $\gamma = \gamma_H = 0.31 \ week^{-1}$;
    \item The two models are indistinguishable by the long term clone size distribution if $\Omega_{DH} = \Omega_H$, that is:
    \begin{equation*}
        \dfrac{r\lambda_H}{\rho_H}  =  \Omega_H  =  \Omega_{DH}  =  \lambda x \dfrac{1+\rho x}{\left(1+x\right)^2}
    \end{equation*}
    
    \noindent Where $x  =  \dfrac{\omega_B}{\rho\lambda}$. Rearranging, we obtained:
    \begin{equation*}
        (\Omega_H \ - \ \lambda\rho)x^2 \ + \ 
        (2\Omega_H \ - \ \lambda)x \ + 
        \Omega_H \ = \ 0
    \end{equation*}
    \noindent This equation is not analytically solvable because $x$ depends on $\lambda$, and it admits real solutions only if $\lambda$ is sufficiently large. \newline 
    We chose the value which is the closest to $\lambda_H$ and for which $x \in \mathbb{R}$. In the end, we set $\lambda  =  1.298 \ week^{-1}$ and obtained $\omega_B  =  0.4295 \ week^{-1}$.
    \item Finally, we computed $\omega_A$ from the balance condition: $\omega_A = \dfrac{\omega_B\gamma}{\lambda}  =  0.1176 \ week^{-1}$. 
\end{itemize}
\subsubsection*{The implementation of the crowding feedback}\label{SEC:3_4_2}

\noindent We studied the effect of a linear crowding feedback on the system's parameters, that is: $\theta(N) \ = \ \theta' \ \pm kN $, where $\theta \ \in \left\{ \lambda, \ \omega_A, \ \omega_B, \ \gamma \right\}$, $k$ was the strength of the feedback and the $\pm$ sign was chosen for each parameter accordingly to the conditions reported above. In addition, we assumed that the feedback affects one parameter at time and $k$ was set to be constant and equal across the parameters, so that the resulting effect could be compared fairly. \newline
Under these assumptions, the feedback on the four parameters was implemented as follows: 

\begin{itemize}
    \item $\ \lambda(N)   =  \lambda'  - kN$
    \item $\omega_A(N)  =  \omega_A' - kN$
    \item $\omega_B(N)  =  \omega_B' + kN$
    \item $\ \gamma(N)    =  \gamma'   + kN$
\end{itemize}

\noindent Given the linear functional form, it is immediate to evaluate the stationary state. Let $\theta_0$ denote the values of the parameters that satisfy eq. (\ref{eq:5}). If $\lambda$ was subject to the feedback, we could find $N^*_{\lambda}$ by imposing $\lambda(N^*) \ = \ \lambda_0$:

\begin{equation*}
    N^*_{\lambda}  =  \dfrac{\lambda'-\lambda_0}{k}
\end{equation*}
\noindent Similar calculations for the other parameters give:
\begin{equation*}
    N^*_{\omega_A}  =  \dfrac{\omega_{A}'-\omega_{A,0}}{k}
    \quad\quad\quad
    N^*_{\omega_B}  =  \dfrac{\omega_{B,0}-\omega_B'}{k}
\end{equation*}
\vspace{-7mm}
\begin{equation*}
    N^*_{\gamma}  =  \dfrac{\gamma_0-\gamma'}{k}
\end{equation*}

\noindent Where $\lambda_0$, $\omega_{A,0}$, $\omega_{B,0}$, $ \gamma_0$ are the values reported in the previous section. To compare the different choices fairly, we imposed the stationary state to be the same across all of them, i.e. $N^*_{\lambda}  =  N^*_{\omega_A}  =  N^*_{\omega_B}  =   N^*_{\gamma}  =  N_{ss}$. \newline
Therefore, the dependence of the four parameters on the $N$ could be rewritten as:

\begin{itemize}
    \item $\ \lambda(N)   = \ \lambda_0     -  k(N  -  N_{ss})$
    \item $\omega_A(N)  =  \omega_{A,0}  -  k(N  -  N_{ss})$
    \item $\omega_B(N)  =  \omega_{B,0}  +  k(N  -  N_{ss})$
    \item $\ \gamma(N)    = \ \gamma_0      +  k(N  -  N_{ss})$
\end{itemize}

\noindent The implementation of the feedback in the model makes any analytical result impossible, especially when it comes to the FPTs. Therefore, we chose to simulate the model by means of Gillespie's first reaction algorithm \cite{gillespie1976general}\cite{erban2007practical}. We carried out $5\cdot10^3$ simulations for each possible choice of the parameter subject to the feedback, for different values of the feedback's strength ($\vert k\vert \in \{ 5\cdot 10^{-4},\ 4.375\cdot 10^{-4},\ 3.375\cdot 10^{-4},\ 3.125\cdot 10^{-4}$, $ 2.5\cdot 10^{-4} \}$), different initial number of cells and stationary state cellular populations.\newline

\begin{figure}[b]
    \includegraphics[width=0.95\linewidth, height = 6cm]{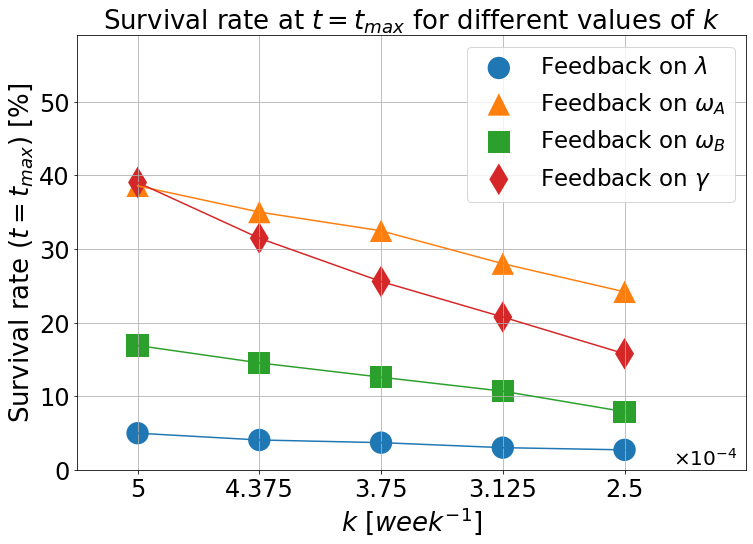}
    \caption{The survival rate of the clones generated by single cells depended on both the parameter subject to it and on the feedback strength. The feedback on $\omega_A$ provided an higher survival probability for all the values of $k$ considered, except for the largest one ($k = 5\cdot 10^{-4}$), for which the survival rate was comparable to that provided by the feedback on $\gamma$.}
    \label{fig:1}
\end{figure}

\section{Results}\label{Results}

\noindent All the results reported below were robust with respect to the specific values chosen for the reaction rates. To verify it, we repeated all the sets of simulations with the parameters $\lambda$, $\omega_A$, $\omega_B$ and $\gamma$ extracted from a Gaussian distribution centered around the values $\theta_0$ reported in Materials and Methods. We imposed that $\theta \sim \mathcal{N}(\theta_0,\theta_0/5)$, where $\theta \in \{\lambda, \omega_B, \gamma \}$, and set $\omega_A$ according to  eq. (\ref{eq:5}). The results of the simulations conducted with the extracted values of the parameters were coherent with those reported below, therefore we could safely assume that our findings did not depend on the parameters' setting procedure adopted, which presented a small degree of arbitrariness.

\subsection*{The feedback heterogeneously affects the survival rate and the steady state population variability}\label{SEC:4_1}

\begin{figure}[b]
    \centering
    \includegraphics[width=0.95\linewidth, height = 6cm]{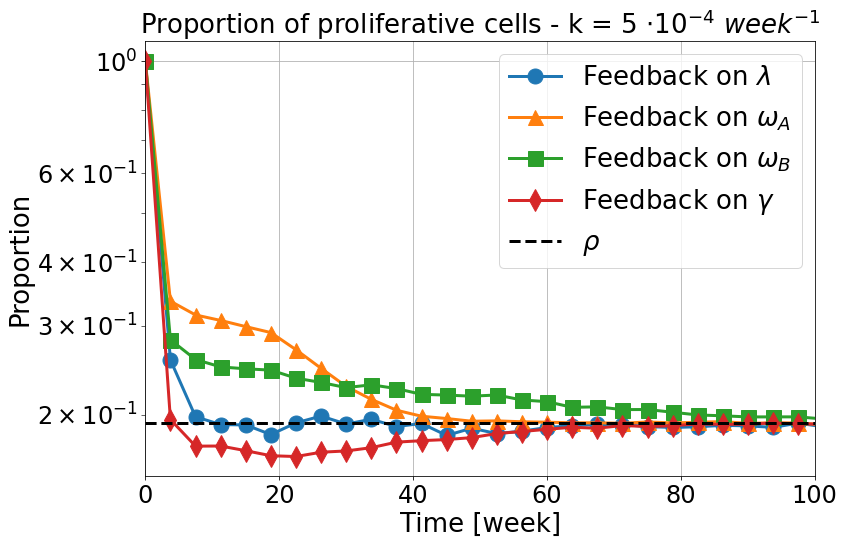}
    \caption{The proportion of proliferative cells as a function of time in growing clones. When the feedback acted on $\omega_A$, there was an initial abundance of proliferative cells, that eventually decayed to the stationary value ($\rho$). In contrast, when the feedback acted on $\gamma$, the early proportion of proliferative cells is smaller than $\rho$, which is attained after weeks.
    }
    \label{fig:2}
\end{figure}

\begin{figure}[t]
    \centering
    \includegraphics[width=0.95\linewidth, height = 6cm]{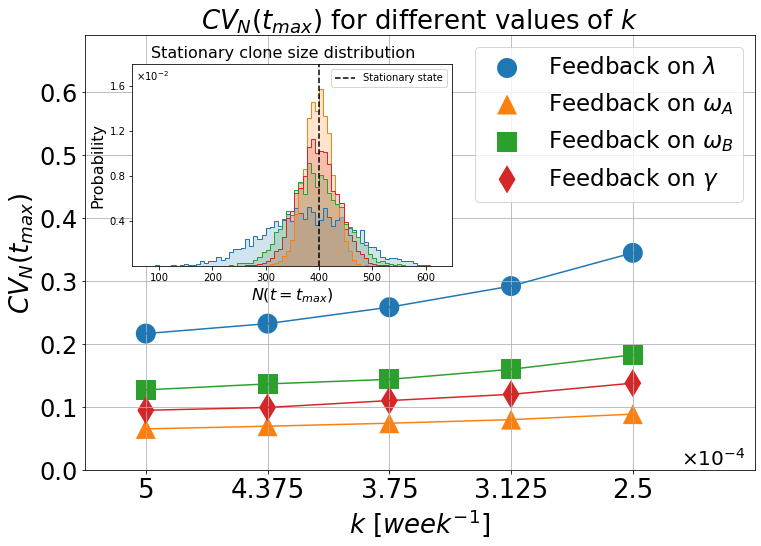}
    \caption{The coefficient of variation of the steady state population $CV_N(t = t_{max})$ strongly depended on the parameter subject to the feedback. For all the values of $k$ considered, the feedback $\omega_A$ and $\gamma$ produced a less variable steady state clone population. \newline
    Box: Stationary probability distributions of the total population ($k=5\cdot 10^{-4}\ week^{-1}$). When the feedback was applied on $\omega_A$ and $\gamma$, the clone size distribution was much more peaked around the mean value ($\approx N_{ss}$) and its deviations from it considerably diminished.}
    \label{fig:3}
\end{figure}

\noindent Critical homeostasis models can also be used to study growth phenomenons (by making the reaction rates unbalanced). Hence, we first studied the scenario in which the simulations start with a single cell and must reach the desired stationary state. \\
We observed that if the simulations start with a single cell, which can be a type A cell with probability $\rho$ or type B cell with probability $1-\rho$, the number of surviving clones does not decay to $0$ as it happens in the model without any feedback \cite{greulich2016dynamic}. Instead, the survival rate reached an asymptote whose value depended on the parameter subject to the feedback and on $k$ (Figure \ref{fig:1}). Nevertheless, the ordering of the survival rates according to the parameter subject to the feedback was invariant with respect to the strength of the feedback. In particular, we observed that the feedbacks on $\omega_A$ provide the highest survival rates for all the values of $k$ considered. All the clones that survive reach the fixed stationary state and, as expected, the stronger the feedback, the higher the survival rate if the parameter subject to the feedback is kept fixed. \\
A measurable quantity in cellular models is the proportion of proliferative cells in the system. We observed that, depending on which parameter the feedback acted on, the fraction of type A cells in the system varied heterogeneously with time (Figure \ref{fig:2}). \\
In particular, the two most efficient feedback mechanisms from the point of view of the survival probability, $\omega_A$ and $\gamma$, adopted opposite strategies: the feedback on $\omega_A$ provided an initially higher percentage of proliferative cells, which eventually decayed to the expected stationary value $\rho$. Instead, the feedback on $\gamma$ resulted in a smaller presence of proliferative cells in the early stages of the dynamics. Anyway, in all the scenarios the steady state proportions were attained eventually.\newline
Afterwards, we focused on the surviving clones, i.e. those that maintain the homeostasis and constitute the steady state population. In particular, we addressed how the presence of the feedback affected the fluctuations of the clones' population around the mean, fixed number. We quantified the magnitude of these fluctuations by mean of the coefficient of variation $CV_N(t_{max})$, defined as the ratio between the standard deviation and the mean value of the total clone population $N$ at time $t = t_{max}$ of the simulations.\\
We found out that the $CV$ varies among the four possible settings of the feedback (Figure \ref{fig:3}), in particular it is smaller when the feedback is applied on $\omega_A$ and $\gamma$, in comparison to the other two possible scenarios, and the difference between them widened as $k$ decreased. \newline
This fact was confirmed by the shape of the steady state distributions of the total population size produced by the simulations. We observed that the feedback on $\omega_A$, and secondly on $\gamma$, provided a distribution much more peaked around the fixed steady state value $N_{ss}$ than the other two feedbacks (Figure \ref{fig:3} - Box).\newline
Therefore, we could deduce that even though all the possible parameter's choices produced a similar, robust stationary state, the effect on the variability of the total clone population is heterogeneous. This fact could render the choice of implementing the feedback on a specific parameter to be more or less favourable from a biological perspective.

\begin{figure}[t]
\centering
\includegraphics[width=\linewidth, height = 6cm]{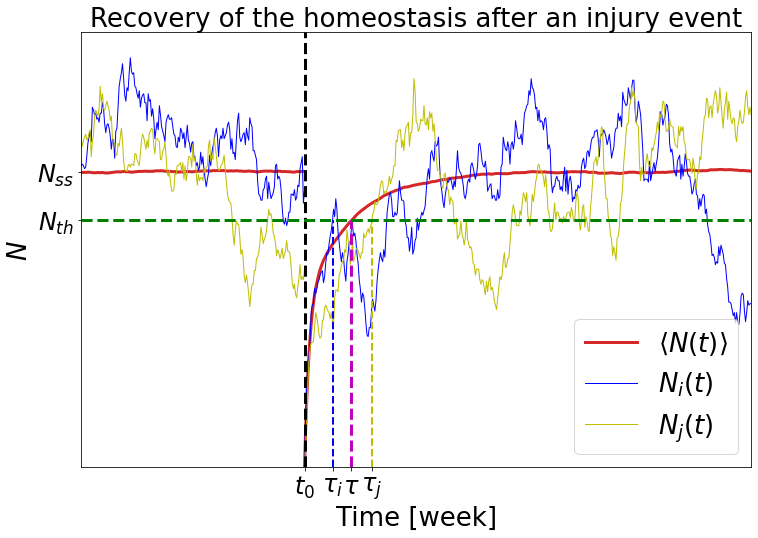}
\caption{Depiction of how the clones recover the homeostatic state following an injury event at time $t_0$. The recovery threshold $N_{th}$ can be thought of the minimum average number of cells per clone necessary for the tissue to regain its functionality. The recovery time $\tau$ is defined as the time it takes the cellular population to reach $N_{th}$. Distinct simulations $i$ and $j$ provide diverse values $\tau_i$ and $\tau_j$, hence we studied whether a specific feedback minimized the average recovery time $\langle \tau \rangle$.}
\label{fig:4}
\end{figure}

\subsection*{Specific choices of the feedback speed up the recovery time of the system}\label{SEC:4_2}

\begin{figure}[t]
    \centering
    \begin{subfigure}[b]{0.49\linewidth}
        \includegraphics[width=\linewidth, height = 6cm]{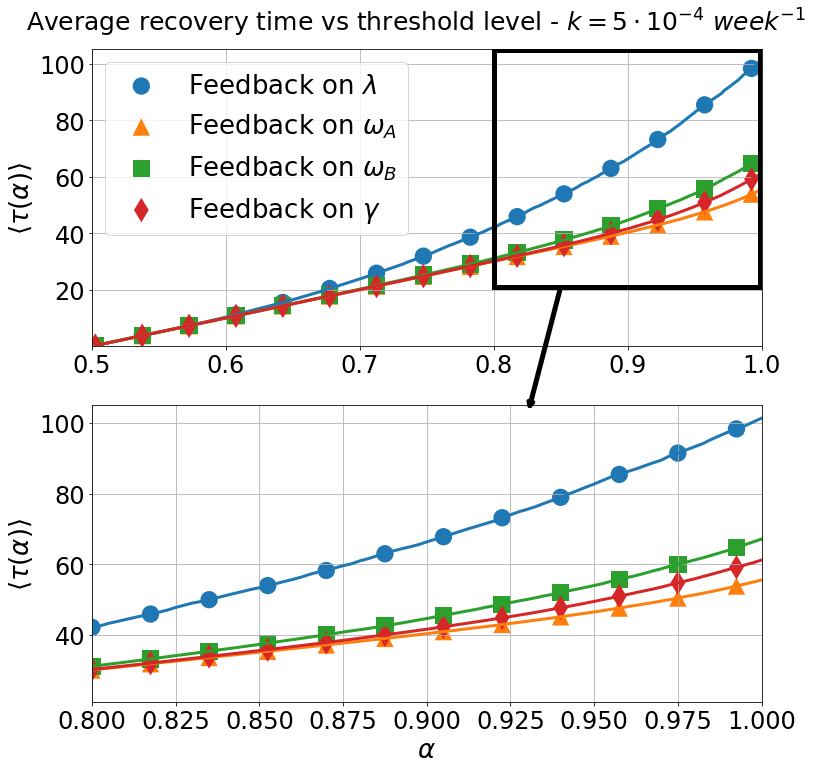}
        \caption{ }
        \label{fig:5A}
    \end{subfigure}
    \hfill
    \begin{subfigure}[b]{0.49\linewidth}
        \includegraphics[width=\linewidth, height = 6cm]{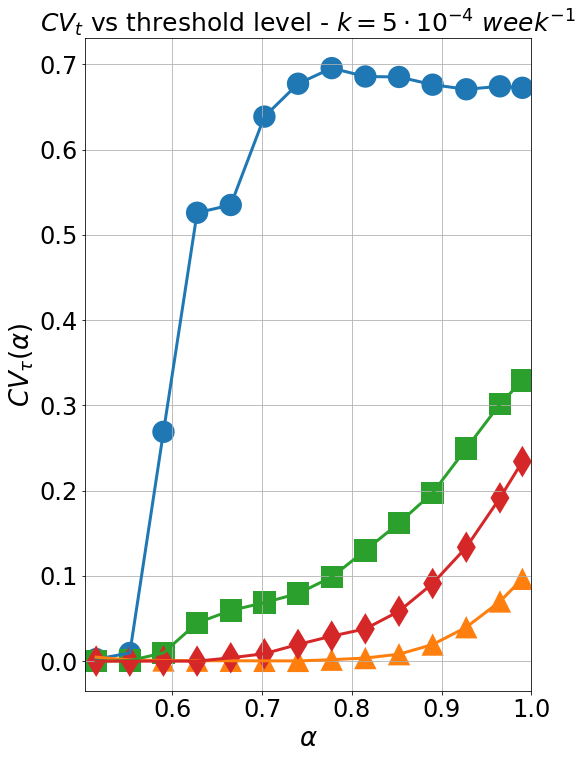}
        \caption{ }
        \label{fig:5B}
    \end{subfigure}
    \caption{(a) Average recovery time $\tau$ as a function of the threshold level $\alpha$, for the four different choices of the parameter subject to the feedback. The lower plot highlights the differences between the feedback applied on the four parameters in the region $\alpha \in (0.8,1)$. \newline
    (b) Coefficient of variation of the recovery time $\tau$ as a function of $\alpha$. In both plots, the strength of the feedback is set to be equal to $k=5\cdot 10^{-4}\ week^{-1}$.}
    \label{fig:5}
\end{figure}

\noindent Finally, we studied the recovery time of the system upon the removal of a macroscopic fraction of cells ($\approx 50\%$ of the homeostatic number). We defined the recovery time $\tau_{\theta}$ as the time it took the cell population of a clone with the feedback applied to $\theta \in \{ \lambda, \ \omega_A, \ \omega_B, \ \gamma \}$ to recover a threshold number $N_{th}$ of cells after the initial loss at time $t_0$ (Figure \ref{fig:4}). Since $N_{th}$ could in principle assume any value $\le N_{ss}$, we studied the recovery time as a function of the ratio $\alpha = N_{th}/N_{ss}$, $\alpha \in \left(N(t_0)/N_{ss},1\right)$.\newline
We found out that the presence of the linear feedback on each parameter resulted in remarkably different outcomes when it came to both the average recovery time $\langle \tau_{\theta}(\alpha) \rangle$  and its variability, which we quantified by means of the coefficient of variation $CV_{\tau_{\theta}}(\alpha)$.\newline
Concerning $\langle \tau_{\theta} \rangle$, the feedback on  $\omega_A$ and $\gamma$ provided a faster average recovery time for all the threshold values considered (Figure \ref{fig:5A} - top). The difference became larger as $\alpha$ increased: in particular, for $\alpha>0.8$, the gap between $\langle \tau_{\omega_A}\rangle$, $\langle \tau_{\gamma} \rangle$ and $\langle \tau_{\omega_B} \rangle$ started to widen (Figure \ref{fig:5A} - bottom), while $\langle \tau_{\lambda} \rangle$ resulted considerably higher for even smaller values of $\alpha$. \newline
When it comes to $CV_{\tau_{\theta}}$, we found out that the feedback implementations that resulted in a faster $\langle \tau_{\theta} \rangle$ also provided a significantly less variable recovery time (Figure \ref{fig:5B}). We found similar results also for the other values of $k$, but the average recovery times and their $CV$s were higher, as expected due to the less intense feedback.

\section{Discussion}

\noindent We studied, by mean of stochastic simulations, an epidermal homeostasis model that takes into account cell's plasticity and the need of a feedback mechanism to maintain the homeostasis. In particular, the feedback has to influence the life cycle and reaction rates of the cells and adjust it to face possible changes in the system. In this regard, we initially focused on the effect the feedback has on the survival probability of a cellular population generated by a single cell, and on the fluctuations of its size at the steady state. \newline 
We found out that the feedback on the differentiated-to-proliferative state switching rate $\omega_A$ provided a better survival probability and a less variable steady state for the clone's population. The second most efficient choice in both points was to apply the feedback on the loss rate of differentiated cells, $\gamma$ \newline
These results suggested that a possibly better strategy to increase the survival chance of the cellular population and decrease its variability, would be to put a control mechanism on the reactions that involve differentiated cells, rather than boosting the frequency of cell divisions or decreasing the differentiation rate of progenitor cells. This can be done by either boosting the state reversal rate in the early stages of the dynamics (\textit{i.e.} increase $\omega_A$) or preventing cells from dying (\textit{i.e.} decrease $\gamma$).\newline
Subsequently, we simulated the recovery of the homeostatic state following the removal of a macroscopic portion of cells, to address whether a specific choice of the parameter made subjected to a linear crowding feedback is preferable from a twofold perspective: the system should repopulate as quickly as possible and the recovery time should not present large fluctuations. \newline
This analysis showed that the feedback on $\omega_A$ and on $\gamma$ was better than the other two choices also in this context: $\langle \tau_{\gamma}\rangle$ and $\langle \tau_{\omega_A} \rangle$ were consistently smaller and the measured clone-to-clone variability was lower compared to the other two parameters, for all the possible threshold values.\newline
Our work showed that, even though any feedback setting makes the model robust to fluctuations of both the reaction rates and the cell population, there are certain quantities that highlight the heterogeneity among the four possible choices and offer a way to discriminate them on an efficiency basis. In particular, the feedback $\omega_A$ and on $\gamma$ constantly outperformed the other two possible implementations in all the quantities we compared them on. \\
In this respect it is interesting to notice (see \cite{gagliardi2021collective,valon2021robustness} for a detailed analysis) that a relevant role in preserving homeostasis in epithelial tissues is played by a feedback control on cell death (i.e. exactly on the parameter $\gamma$ discussed above) mediated by the ERK/Akt pathway which is triggered by apoptotic cells and induces survival of neighboring cells.\newline
This example, and the recent observation that epithelial cells are affected by the activity
of their neighbors and tend to respond to external perturbations in a spatio-temporal coordinated way  \cite{mesa2018homeostatic,rompolas2016spatiotemporal}  show the importance of keeping into account the spatial information (and in particular nearest-neighbour interactions) in modelling homeostasis. This issue is outside the scope of the present paper but it should certainly deserve further attention and we plan to address it in a forthcoming publication.\newline
In principle, the cells could present some control mechanism on all its physiological parameters, and not only on one of them as we studied in this work. However, the maintenance of a feedback mechanism is costly for the cells, hence the economic principles behind their physiology may prompt the adoption of the most efficient mechanism to face stressful situations.\newline
Further experiments and measurements should be carried out to shed light on the real effect and functioning of feedback mechanisms in our cells in a more quantitative and detailed way. For instance, one could measure the proportion of proliferative cells during the growth or the recovery of the tissue to check if the experiments were compatible with our simulated results (Figure \ref{fig:2}). \newline
In addition to these open questions, also the specific functional form of the feedback mechanism is yet to be determined. We focused on a linear feedback only for the sake of simplicity, however other functional forms may be more suitable, efficient or realistic. For instance, it has been observed that a model involving both linear and cooperative feedbacks is compatible with experimental data of the immune system \cite{hart2012design}\cite{hart2014paradoxical}, hence future studies may address the implementation of other forms of feedbacks that resemble more closely the actual biology of the epidermis.

\bibliography{Bibliography}

\begin{thebibliography}{49}%
\makeatletter
\providecommand \@ifxundefined [1]{%
 \@ifx{#1\undefined}
}%
\providecommand \@ifnum [1]{%
 \ifnum #1\expandafter \@firstoftwo
 \else \expandafter \@secondoftwo
 \fi
}%
\providecommand \@ifx [1]{%
 \ifx #1\expandafter \@firstoftwo
 \else \expandafter \@secondoftwo
 \fi
}%
\providecommand \natexlab [1]{#1}%
\providecommand \enquote  [1]{``#1''}%
\providecommand \bibnamefont  [1]{#1}%
\providecommand \bibfnamefont [1]{#1}%
\providecommand \citenamefont [1]{#1}%
\providecommand \href@noop [0]{\@secondoftwo}%
\providecommand \href [0]{\begingroup \@sanitize@url \@href}%
\providecommand \@href[1]{\@@startlink{#1}\@@href}%
\providecommand \@@href[1]{\endgroup#1\@@endlink}%
\providecommand \@sanitize@url [0]{\catcode `\\12\catcode `\$12\catcode
  `\&12\catcode `\#12\catcode `\^12\catcode `\_12\catcode `\%12\relax}%
\providecommand \@@startlink[1]{}%
\providecommand \@@endlink[0]{}%
\providecommand \url  [0]{\begingroup\@sanitize@url \@url }%
\providecommand \@url [1]{\endgroup\@href {#1}{\urlprefix }}%
\providecommand \urlprefix  [0]{URL }%
\providecommand \Eprint [0]{\href }%
\providecommand \doibase [0]{https://doi.org/}%
\providecommand \selectlanguage [0]{\@gobble}%
\providecommand \bibinfo  [0]{\@secondoftwo}%
\providecommand \bibfield  [0]{\@secondoftwo}%
\providecommand \translation [1]{[#1]}%
\providecommand \BibitemOpen [0]{}%
\providecommand \bibitemStop [0]{}%
\providecommand \bibitemNoStop [0]{.\EOS\space}%
\providecommand \EOS [0]{\spacefactor3000\relax}%
\providecommand \BibitemShut  [1]{\csname bibitem#1\endcsname}%
\let\auto@bib@innerbib\@empty
\bibitem [{\citenamefont {Antal}\ and\ \citenamefont
  {Krapivsky}(2010)}]{antal2010exact}%
  \BibitemOpen
  \bibfield  {author} {\bibinfo {author} {\bibfnamefont {T.}~\bibnamefont
  {Antal}}\ and\ \bibinfo {author} {\bibfnamefont {P.}~\bibnamefont
  {Krapivsky}},\ }\href@noop {} {\bibfield  {journal} {\bibinfo  {journal}
  {Journal of Statistical Mechanics: Theory and Experiment}\ }\textbf {\bibinfo
  {volume} {2010}},\ \bibinfo {pages} {P07028} (\bibinfo {year}
  {2010})}\BibitemShut {NoStop}%
\bibitem [{\citenamefont {McGrath}\ \emph {et~al.}(2004)\citenamefont
  {McGrath}, \citenamefont {Eady},\ and\ \citenamefont
  {Pope}}]{mcgrath2004anatomy}%
  \BibitemOpen
  \bibfield  {author} {\bibinfo {author} {\bibfnamefont {J.}~\bibnamefont
  {McGrath}}, \bibinfo {author} {\bibfnamefont {R.}~\bibnamefont {Eady}},\ and\
  \bibinfo {author} {\bibfnamefont {F.}~\bibnamefont {Pope}},\ }\href@noop {}
  {\bibfield  {journal} {\bibinfo  {journal} {Rook’s textbook of
  dermatology}\ }\textbf {\bibinfo {volume} {1}},\ \bibinfo {pages} {3}
  (\bibinfo {year} {2004})}\BibitemShut {NoStop}%
\bibitem [{\citenamefont {Augustin}(2015)}]{augustin2015wnt}%
  \BibitemOpen
  \bibfield  {author} {\bibinfo {author} {\bibfnamefont {I.}~\bibnamefont
  {Augustin}},\ }\href@noop {} {\bibfield  {journal} {\bibinfo  {journal}
  {JDDG: Journal der Deutschen Dermatologischen Gesellschaft}\ }\textbf
  {\bibinfo {volume} {13}},\ \bibinfo {pages} {302} (\bibinfo {year}
  {2015})}\BibitemShut {NoStop}%
\bibitem [{\citenamefont {Garg}\ \emph {et~al.}(2001)\citenamefont {Garg},
  \citenamefont {Chren}, \citenamefont {Sands}, \citenamefont {Matsui},
  \citenamefont {Marenus}, \citenamefont {Feingold},\ and\ \citenamefont
  {Elias}}]{garg2001psychological}%
  \BibitemOpen
  \bibfield  {author} {\bibinfo {author} {\bibfnamefont {A.}~\bibnamefont
  {Garg}}, \bibinfo {author} {\bibfnamefont {M.-M.}\ \bibnamefont {Chren}},
  \bibinfo {author} {\bibfnamefont {L.~P.}\ \bibnamefont {Sands}}, \bibinfo
  {author} {\bibfnamefont {M.~S.}\ \bibnamefont {Matsui}}, \bibinfo {author}
  {\bibfnamefont {K.~D.}\ \bibnamefont {Marenus}}, \bibinfo {author}
  {\bibfnamefont {K.~R.}\ \bibnamefont {Feingold}},\ and\ \bibinfo {author}
  {\bibfnamefont {P.~M.}\ \bibnamefont {Elias}},\ }\href@noop {} {\bibfield
  {journal} {\bibinfo  {journal} {Archives of dermatology}\ }\textbf {\bibinfo
  {volume} {137}},\ \bibinfo {pages} {53} (\bibinfo {year} {2001})}\BibitemShut
  {NoStop}%
\bibitem [{\citenamefont {Potten}(1996)}]{potten1996stem}%
  \BibitemOpen
  \bibfield  {author} {\bibinfo {author} {\bibfnamefont {C.}~\bibnamefont
  {Potten}},\ }\href@noop {} {\emph {\bibinfo {title} {Stem cells}}}\ (\bibinfo
   {publisher} {Elsevier},\ \bibinfo {year} {1996})\BibitemShut {NoStop}%
\bibitem [{\citenamefont {Allen}\ and\ \citenamefont
  {Potten}(1974)}]{allen1974fine}%
  \BibitemOpen
  \bibfield  {author} {\bibinfo {author} {\bibfnamefont {T.~D.}\ \bibnamefont
  {Allen}}\ and\ \bibinfo {author} {\bibfnamefont {C.~S.}\ \bibnamefont
  {Potten}},\ }\href@noop {} {\bibfield  {journal} {\bibinfo  {journal}
  {Journal of cell science}\ }\textbf {\bibinfo {volume} {15}},\ \bibinfo
  {pages} {291} (\bibinfo {year} {1974})}\BibitemShut {NoStop}%
\bibitem [{\citenamefont {Potten}(1981)}]{potten1981cell}%
  \BibitemOpen
  \bibfield  {author} {\bibinfo {author} {\bibfnamefont {C.}~\bibnamefont
  {Potten}},\ }\href@noop {} {\bibfield  {journal} {\bibinfo  {journal}
  {International review of cytology}\ }\textbf {\bibinfo {volume} {69}},\
  \bibinfo {pages} {271} (\bibinfo {year} {1981})}\BibitemShut {NoStop}%
\bibitem [{\citenamefont {Kretzschmar}\ and\ \citenamefont
  {Watt}(2012)}]{kretzschmar2012lineage}%
  \BibitemOpen
  \bibfield  {author} {\bibinfo {author} {\bibfnamefont {K.}~\bibnamefont
  {Kretzschmar}}\ and\ \bibinfo {author} {\bibfnamefont {F.~M.}\ \bibnamefont
  {Watt}},\ }\href@noop {} {\bibfield  {journal} {\bibinfo  {journal} {Cell}\
  }\textbf {\bibinfo {volume} {148}},\ \bibinfo {pages} {33} (\bibinfo {year}
  {2012})}\BibitemShut {NoStop}%
\bibitem [{\citenamefont {Clayton}\ \emph {et~al.}(2007)\citenamefont
  {Clayton}, \citenamefont {Doup{\'e}}, \citenamefont {Klein}, \citenamefont
  {Winton}, \citenamefont {Simons},\ and\ \citenamefont
  {Jones}}]{clayton2007single}%
  \BibitemOpen
  \bibfield  {author} {\bibinfo {author} {\bibfnamefont {E.}~\bibnamefont
  {Clayton}}, \bibinfo {author} {\bibfnamefont {D.~P.}\ \bibnamefont
  {Doup{\'e}}}, \bibinfo {author} {\bibfnamefont {A.~M.}\ \bibnamefont
  {Klein}}, \bibinfo {author} {\bibfnamefont {D.~J.}\ \bibnamefont {Winton}},
  \bibinfo {author} {\bibfnamefont {B.~D.}\ \bibnamefont {Simons}},\ and\
  \bibinfo {author} {\bibfnamefont {P.~H.}\ \bibnamefont {Jones}},\ }\href@noop
  {} {\bibfield  {journal} {\bibinfo  {journal} {Nature}\ }\textbf {\bibinfo
  {volume} {446}},\ \bibinfo {pages} {185} (\bibinfo {year}
  {2007})}\BibitemShut {NoStop}%
\bibitem [{\citenamefont {Dekoninck}\ \emph {et~al.}(2020)\citenamefont
  {Dekoninck}, \citenamefont {Hannezo}, \citenamefont {Sifrim}, \citenamefont
  {Miroshnikova}, \citenamefont {Aragona}, \citenamefont {Malfait},
  \citenamefont {Gargouri}, \citenamefont {De~Neunheuser}, \citenamefont
  {Dubois}, \citenamefont {Voet} \emph {et~al.}}]{dekoninck2020defining}%
  \BibitemOpen
  \bibfield  {author} {\bibinfo {author} {\bibfnamefont {S.}~\bibnamefont
  {Dekoninck}}, \bibinfo {author} {\bibfnamefont {E.}~\bibnamefont {Hannezo}},
  \bibinfo {author} {\bibfnamefont {A.}~\bibnamefont {Sifrim}}, \bibinfo
  {author} {\bibfnamefont {Y.~A.}\ \bibnamefont {Miroshnikova}}, \bibinfo
  {author} {\bibfnamefont {M.}~\bibnamefont {Aragona}}, \bibinfo {author}
  {\bibfnamefont {M.}~\bibnamefont {Malfait}}, \bibinfo {author} {\bibfnamefont
  {S.}~\bibnamefont {Gargouri}}, \bibinfo {author} {\bibfnamefont
  {C.}~\bibnamefont {De~Neunheuser}}, \bibinfo {author} {\bibfnamefont
  {C.}~\bibnamefont {Dubois}}, \bibinfo {author} {\bibfnamefont
  {T.}~\bibnamefont {Voet}}, \emph {et~al.},\ }\href@noop {} {\bibfield
  {journal} {\bibinfo  {journal} {Cell}\ }\textbf {\bibinfo {volume} {181}},\
  \bibinfo {pages} {604} (\bibinfo {year} {2020})}\BibitemShut {NoStop}%
\bibitem [{\citenamefont {Mascr{\'e}}\ \emph {et~al.}(2012)\citenamefont
  {Mascr{\'e}}, \citenamefont {Dekoninck}, \citenamefont {Drogat},
  \citenamefont {Youssef}, \citenamefont {Broh{\'e}e}, \citenamefont
  {Sotiropoulou}, \citenamefont {Simons},\ and\ \citenamefont
  {Blanpain}}]{mascre2012distinct}%
  \BibitemOpen
  \bibfield  {author} {\bibinfo {author} {\bibfnamefont {G.}~\bibnamefont
  {Mascr{\'e}}}, \bibinfo {author} {\bibfnamefont {S.}~\bibnamefont
  {Dekoninck}}, \bibinfo {author} {\bibfnamefont {B.}~\bibnamefont {Drogat}},
  \bibinfo {author} {\bibfnamefont {K.~K.}\ \bibnamefont {Youssef}}, \bibinfo
  {author} {\bibfnamefont {S.}~\bibnamefont {Broh{\'e}e}}, \bibinfo {author}
  {\bibfnamefont {P.~A.}\ \bibnamefont {Sotiropoulou}}, \bibinfo {author}
  {\bibfnamefont {B.~D.}\ \bibnamefont {Simons}},\ and\ \bibinfo {author}
  {\bibfnamefont {C.}~\bibnamefont {Blanpain}},\ }\href@noop {} {\bibfield
  {journal} {\bibinfo  {journal} {Nature}\ }\textbf {\bibinfo {volume} {489}},\
  \bibinfo {pages} {257} (\bibinfo {year} {2012})}\BibitemShut {NoStop}%
\bibitem [{\citenamefont {Lei}(2020)}]{lei2020general}%
  \BibitemOpen
  \bibfield  {author} {\bibinfo {author} {\bibfnamefont {J.}~\bibnamefont
  {Lei}},\ }\href@noop {} {\bibfield  {journal} {\bibinfo  {journal} {Journal
  of Theoretical Biology}\ }\textbf {\bibinfo {volume} {492}},\ \bibinfo
  {pages} {110196} (\bibinfo {year} {2020})}\BibitemShut {NoStop}%
\bibitem [{\citenamefont {Blanpain}\ and\ \citenamefont
  {Fuchs}(2014)}]{blanpain2014plasticity}%
  \BibitemOpen
  \bibfield  {author} {\bibinfo {author} {\bibfnamefont {C.}~\bibnamefont
  {Blanpain}}\ and\ \bibinfo {author} {\bibfnamefont {E.}~\bibnamefont
  {Fuchs}},\ }\href@noop {} {\bibfield  {journal} {\bibinfo  {journal}
  {Science}\ }\textbf {\bibinfo {volume} {344}} (\bibinfo {year}
  {2014})}\BibitemShut {NoStop}%
\bibitem [{\citenamefont {Doup{\'e}}\ \emph {et~al.}(2012)\citenamefont
  {Doup{\'e}}, \citenamefont {Alcolea}, \citenamefont {Roshan}, \citenamefont
  {Zhang}, \citenamefont {Klein}, \citenamefont {Simons},\ and\ \citenamefont
  {Jones}}]{doupe2012single}%
  \BibitemOpen
  \bibfield  {author} {\bibinfo {author} {\bibfnamefont {D.~P.}\ \bibnamefont
  {Doup{\'e}}}, \bibinfo {author} {\bibfnamefont {M.~P.}\ \bibnamefont
  {Alcolea}}, \bibinfo {author} {\bibfnamefont {A.}~\bibnamefont {Roshan}},
  \bibinfo {author} {\bibfnamefont {G.}~\bibnamefont {Zhang}}, \bibinfo
  {author} {\bibfnamefont {A.~M.}\ \bibnamefont {Klein}}, \bibinfo {author}
  {\bibfnamefont {B.~D.}\ \bibnamefont {Simons}},\ and\ \bibinfo {author}
  {\bibfnamefont {P.~H.}\ \bibnamefont {Jones}},\ }\href@noop {} {\bibfield
  {journal} {\bibinfo  {journal} {Science}\ }\textbf {\bibinfo {volume}
  {337}},\ \bibinfo {pages} {1091} (\bibinfo {year} {2012})}\BibitemShut
  {NoStop}%
\bibitem [{\citenamefont {Ritsma}\ \emph {et~al.}(2014)\citenamefont {Ritsma},
  \citenamefont {Ellenbroek}, \citenamefont {Zomer}, \citenamefont {Snippert},
  \citenamefont {de~Sauvage}, \citenamefont {Simons}, \citenamefont {Clevers},\
  and\ \citenamefont {van Rheenen}}]{ritsma2014intestinal}%
  \BibitemOpen
  \bibfield  {author} {\bibinfo {author} {\bibfnamefont {L.}~\bibnamefont
  {Ritsma}}, \bibinfo {author} {\bibfnamefont {S.~I.}\ \bibnamefont
  {Ellenbroek}}, \bibinfo {author} {\bibfnamefont {A.}~\bibnamefont {Zomer}},
  \bibinfo {author} {\bibfnamefont {H.~J.}\ \bibnamefont {Snippert}}, \bibinfo
  {author} {\bibfnamefont {F.~J.}\ \bibnamefont {de~Sauvage}}, \bibinfo
  {author} {\bibfnamefont {B.~D.}\ \bibnamefont {Simons}}, \bibinfo {author}
  {\bibfnamefont {H.}~\bibnamefont {Clevers}},\ and\ \bibinfo {author}
  {\bibfnamefont {J.}~\bibnamefont {van Rheenen}},\ }\href@noop {} {\bibfield
  {journal} {\bibinfo  {journal} {Nature}\ }\textbf {\bibinfo {volume} {507}},\
  \bibinfo {pages} {362} (\bibinfo {year} {2014})}\BibitemShut {NoStop}%
\bibitem [{\citenamefont {Hara}\ \emph {et~al.}(2014)\citenamefont {Hara},
  \citenamefont {Nakagawa}, \citenamefont {Enomoto}, \citenamefont {Suzuki},
  \citenamefont {Yamamoto}, \citenamefont {Simons},\ and\ \citenamefont
  {Yoshida}}]{hara2014mouse}%
  \BibitemOpen
  \bibfield  {author} {\bibinfo {author} {\bibfnamefont {K.}~\bibnamefont
  {Hara}}, \bibinfo {author} {\bibfnamefont {T.}~\bibnamefont {Nakagawa}},
  \bibinfo {author} {\bibfnamefont {H.}~\bibnamefont {Enomoto}}, \bibinfo
  {author} {\bibfnamefont {M.}~\bibnamefont {Suzuki}}, \bibinfo {author}
  {\bibfnamefont {M.}~\bibnamefont {Yamamoto}}, \bibinfo {author}
  {\bibfnamefont {B.~D.}\ \bibnamefont {Simons}},\ and\ \bibinfo {author}
  {\bibfnamefont {S.}~\bibnamefont {Yoshida}},\ }\href@noop {} {\bibfield
  {journal} {\bibinfo  {journal} {Cell stem cell}\ }\textbf {\bibinfo {volume}
  {14}},\ \bibinfo {pages} {658} (\bibinfo {year} {2014})}\BibitemShut
  {NoStop}%
\bibitem [{\citenamefont {Kai}\ and\ \citenamefont
  {Spradling}(2004)}]{kai2004differentiating}%
  \BibitemOpen
  \bibfield  {author} {\bibinfo {author} {\bibfnamefont {T.}~\bibnamefont
  {Kai}}\ and\ \bibinfo {author} {\bibfnamefont {A.}~\bibnamefont
  {Spradling}},\ }\href@noop {} {\bibfield  {journal} {\bibinfo  {journal}
  {Nature}\ }\textbf {\bibinfo {volume} {428}},\ \bibinfo {pages} {564}
  (\bibinfo {year} {2004})}\BibitemShut {NoStop}%
\bibitem [{\citenamefont {Donati}\ and\ \citenamefont
  {Watt}(2015)}]{donati2015stem}%
  \BibitemOpen
  \bibfield  {author} {\bibinfo {author} {\bibfnamefont {G.}~\bibnamefont
  {Donati}}\ and\ \bibinfo {author} {\bibfnamefont {F.~M.}\ \bibnamefont
  {Watt}},\ }\href@noop {} {\bibfield  {journal} {\bibinfo  {journal} {Cell
  stem cell}\ }\textbf {\bibinfo {volume} {16}},\ \bibinfo {pages} {465}
  (\bibinfo {year} {2015})}\BibitemShut {NoStop}%
\bibitem [{\citenamefont {Alcolea}\ \emph {et~al.}(2014)\citenamefont
  {Alcolea}, \citenamefont {Greulich}, \citenamefont {Wabik}, \citenamefont
  {Frede}, \citenamefont {Simons},\ and\ \citenamefont
  {Jones}}]{alcolea2014differentiation}%
  \BibitemOpen
  \bibfield  {author} {\bibinfo {author} {\bibfnamefont {M.~P.}\ \bibnamefont
  {Alcolea}}, \bibinfo {author} {\bibfnamefont {P.}~\bibnamefont {Greulich}},
  \bibinfo {author} {\bibfnamefont {A.}~\bibnamefont {Wabik}}, \bibinfo
  {author} {\bibfnamefont {J.}~\bibnamefont {Frede}}, \bibinfo {author}
  {\bibfnamefont {B.~D.}\ \bibnamefont {Simons}},\ and\ \bibinfo {author}
  {\bibfnamefont {P.~H.}\ \bibnamefont {Jones}},\ }\href@noop {} {\bibfield
  {journal} {\bibinfo  {journal} {Nature cell biology}\ }\textbf {\bibinfo
  {volume} {16}},\ \bibinfo {pages} {612} (\bibinfo {year} {2014})}\BibitemShut
  {NoStop}%
\bibitem [{\citenamefont {Morgan}\ \emph {et~al.}(2018)\citenamefont {Morgan},
  \citenamefont {Mortensson},\ and\ \citenamefont
  {Williams}}]{morgan2018targeting}%
  \BibitemOpen
  \bibfield  {author} {\bibinfo {author} {\bibfnamefont {R.}~\bibnamefont
  {Morgan}}, \bibinfo {author} {\bibfnamefont {E.}~\bibnamefont {Mortensson}},\
  and\ \bibinfo {author} {\bibfnamefont {A.}~\bibnamefont {Williams}},\
  }\href@noop {} {\bibfield  {journal} {\bibinfo  {journal} {British journal of
  cancer}\ }\textbf {\bibinfo {volume} {118}},\ \bibinfo {pages} {1410}
  (\bibinfo {year} {2018})}\BibitemShut {NoStop}%
\bibitem [{\citenamefont {Shimokawa}\ \emph {et~al.}(2017)\citenamefont
  {Shimokawa}, \citenamefont {Ohta}, \citenamefont {Nishikori}, \citenamefont
  {Matano}, \citenamefont {Takano}, \citenamefont {Fujii}, \citenamefont
  {Sugimoto}, \citenamefont {Kanai}, \citenamefont {Sato} \emph
  {et~al.}}]{shimokawa2017visualization}%
  \BibitemOpen
  \bibfield  {author} {\bibinfo {author} {\bibfnamefont {M.}~\bibnamefont
  {Shimokawa}}, \bibinfo {author} {\bibfnamefont {Y.}~\bibnamefont {Ohta}},
  \bibinfo {author} {\bibfnamefont {S.}~\bibnamefont {Nishikori}}, \bibinfo
  {author} {\bibfnamefont {M.}~\bibnamefont {Matano}}, \bibinfo {author}
  {\bibfnamefont {A.}~\bibnamefont {Takano}}, \bibinfo {author} {\bibfnamefont
  {M.}~\bibnamefont {Fujii}}, \bibinfo {author} {\bibfnamefont
  {S.}~\bibnamefont {Sugimoto}}, \bibinfo {author} {\bibfnamefont
  {T.}~\bibnamefont {Kanai}}, \bibinfo {author} {\bibfnamefont
  {T.}~\bibnamefont {Sato}}, \emph {et~al.},\ }\href@noop {} {\bibfield
  {journal} {\bibinfo  {journal} {Nature}\ }\textbf {\bibinfo {volume} {545}},\
  \bibinfo {pages} {187} (\bibinfo {year} {2017})}\BibitemShut {NoStop}%
\bibitem [{\citenamefont {Tata}\ \emph {et~al.}(2013)\citenamefont {Tata},
  \citenamefont {Mou}, \citenamefont {Pardo-Saganta}, \citenamefont {Zhao},
  \citenamefont {Prabhu}, \citenamefont {Law}, \citenamefont {Vinarsky},
  \citenamefont {Cho}, \citenamefont {Breton}, \citenamefont {Sahay} \emph
  {et~al.}}]{tata2013dedifferentiation}%
  \BibitemOpen
  \bibfield  {author} {\bibinfo {author} {\bibfnamefont {P.~R.}\ \bibnamefont
  {Tata}}, \bibinfo {author} {\bibfnamefont {H.}~\bibnamefont {Mou}}, \bibinfo
  {author} {\bibfnamefont {A.}~\bibnamefont {Pardo-Saganta}}, \bibinfo {author}
  {\bibfnamefont {R.}~\bibnamefont {Zhao}}, \bibinfo {author} {\bibfnamefont
  {M.}~\bibnamefont {Prabhu}}, \bibinfo {author} {\bibfnamefont {B.~M.}\
  \bibnamefont {Law}}, \bibinfo {author} {\bibfnamefont {V.}~\bibnamefont
  {Vinarsky}}, \bibinfo {author} {\bibfnamefont {J.~L.}\ \bibnamefont {Cho}},
  \bibinfo {author} {\bibfnamefont {S.}~\bibnamefont {Breton}}, \bibinfo
  {author} {\bibfnamefont {A.}~\bibnamefont {Sahay}}, \emph {et~al.},\
  }\href@noop {} {\bibfield  {journal} {\bibinfo  {journal} {Nature}\ }\textbf
  {\bibinfo {volume} {503}},\ \bibinfo {pages} {218} (\bibinfo {year}
  {2013})}\BibitemShut {NoStop}%
\bibitem [{\citenamefont {Roshan}\ \emph {et~al.}(2016)\citenamefont {Roshan},
  \citenamefont {Murai}, \citenamefont {Fowler}, \citenamefont {Simons},
  \citenamefont {Nikolaidou-Neokosmidou},\ and\ \citenamefont
  {Jones}}]{roshan2016human}%
  \BibitemOpen
  \bibfield  {author} {\bibinfo {author} {\bibfnamefont {A.}~\bibnamefont
  {Roshan}}, \bibinfo {author} {\bibfnamefont {K.}~\bibnamefont {Murai}},
  \bibinfo {author} {\bibfnamefont {J.}~\bibnamefont {Fowler}}, \bibinfo
  {author} {\bibfnamefont {B.~D.}\ \bibnamefont {Simons}}, \bibinfo {author}
  {\bibfnamefont {V.}~\bibnamefont {Nikolaidou-Neokosmidou}},\ and\ \bibinfo
  {author} {\bibfnamefont {P.~H.}\ \bibnamefont {Jones}},\ }\href@noop {}
  {\bibfield  {journal} {\bibinfo  {journal} {Nature cell biology}\ }\textbf
  {\bibinfo {volume} {18}},\ \bibinfo {pages} {145} (\bibinfo {year}
  {2016})}\BibitemShut {NoStop}%
\bibitem [{\citenamefont {Lander}\ \emph {et~al.}(2009)\citenamefont {Lander},
  \citenamefont {Gokoffski}, \citenamefont {Wan}, \citenamefont {Nie},\ and\
  \citenamefont {Calof}}]{lander2009cell}%
  \BibitemOpen
  \bibfield  {author} {\bibinfo {author} {\bibfnamefont {A.~D.}\ \bibnamefont
  {Lander}}, \bibinfo {author} {\bibfnamefont {K.~K.}\ \bibnamefont
  {Gokoffski}}, \bibinfo {author} {\bibfnamefont {F.~Y.~M.}\ \bibnamefont
  {Wan}}, \bibinfo {author} {\bibfnamefont {Q.}~\bibnamefont {Nie}},\ and\
  \bibinfo {author} {\bibfnamefont {A.~L.}\ \bibnamefont {Calof}},\ }\href@noop
  {} {\bibfield  {journal} {\bibinfo  {journal} {PLoS biology}\ }\textbf
  {\bibinfo {volume} {7}},\ \bibinfo {pages} {e1000015} (\bibinfo {year}
  {2009})}\BibitemShut {NoStop}%
\bibitem [{\citenamefont {Asmussen}\ \emph {et~al.}(1983)\citenamefont
  {Asmussen}, \citenamefont {Hering} \emph {et~al.}}]{asmussen1983branching}%
  \BibitemOpen
  \bibfield  {author} {\bibinfo {author} {\bibfnamefont {S.}~\bibnamefont
  {Asmussen}}, \bibinfo {author} {\bibfnamefont {H.}~\bibnamefont {Hering}},
  \emph {et~al.},\ }\href@noop {} {\emph {\bibinfo {title} {Branching
  processes}}},\ Vol.~\bibinfo {volume} {3}\ (\bibinfo  {publisher}
  {Springer},\ \bibinfo {year} {1983})\BibitemShut {NoStop}%
\bibitem [{\citenamefont {Strogatz}(2018)}]{strogatz2018nonlinear}%
  \BibitemOpen
  \bibfield  {author} {\bibinfo {author} {\bibfnamefont {S.~H.}\ \bibnamefont
  {Strogatz}},\ }\href@noop {} {\emph {\bibinfo {title} {Nonlinear dynamics and
  chaos with student solutions manual: With applications to physics, biology,
  chemistry, and engineering}}}\ (\bibinfo  {publisher} {CRC press},\ \bibinfo
  {year} {2018})\BibitemShut {NoStop}%
\bibitem [{\citenamefont {Shraiman}(2005)}]{shraiman2005mechanical}%
  \BibitemOpen
  \bibfield  {author} {\bibinfo {author} {\bibfnamefont {B.~I.}\ \bibnamefont
  {Shraiman}},\ }\href@noop {} {\bibfield  {journal} {\bibinfo  {journal}
  {Proceedings of the National Academy of Sciences}\ }\textbf {\bibinfo
  {volume} {102}},\ \bibinfo {pages} {3318} (\bibinfo {year}
  {2005})}\BibitemShut {NoStop}%
\bibitem [{\citenamefont {Alric}\ \emph {et~al.}(2022)\citenamefont {Alric},
  \citenamefont {Formosa-Dague}, \citenamefont {Dague}, \citenamefont {Holt},\
  and\ \citenamefont {Delarue}}]{alric2022macromolecular}%
  \BibitemOpen
  \bibfield  {author} {\bibinfo {author} {\bibfnamefont {B.}~\bibnamefont
  {Alric}}, \bibinfo {author} {\bibfnamefont {C.}~\bibnamefont
  {Formosa-Dague}}, \bibinfo {author} {\bibfnamefont {E.}~\bibnamefont
  {Dague}}, \bibinfo {author} {\bibfnamefont {L.~J.}\ \bibnamefont {Holt}},\
  and\ \bibinfo {author} {\bibfnamefont {M.}~\bibnamefont {Delarue}},\
  }\href@noop {} {\bibfield  {journal} {\bibinfo  {journal} {Nature Physics}\
  ,\ \bibinfo {pages} {1}} (\bibinfo {year} {2022})}\BibitemShut {NoStop}%
\bibitem [{\citenamefont {Puliafito}\ \emph {et~al.}(2012)\citenamefont
  {Puliafito}, \citenamefont {Hufnagel}, \citenamefont {Neveu}, \citenamefont
  {Streichan}, \citenamefont {Sigal}, \citenamefont {Fygenson},\ and\
  \citenamefont {Shraiman}}]{puliafito2012collective}%
  \BibitemOpen
  \bibfield  {author} {\bibinfo {author} {\bibfnamefont {A.}~\bibnamefont
  {Puliafito}}, \bibinfo {author} {\bibfnamefont {L.}~\bibnamefont {Hufnagel}},
  \bibinfo {author} {\bibfnamefont {P.}~\bibnamefont {Neveu}}, \bibinfo
  {author} {\bibfnamefont {S.}~\bibnamefont {Streichan}}, \bibinfo {author}
  {\bibfnamefont {A.}~\bibnamefont {Sigal}}, \bibinfo {author} {\bibfnamefont
  {D.~K.}\ \bibnamefont {Fygenson}},\ and\ \bibinfo {author} {\bibfnamefont
  {B.~I.}\ \bibnamefont {Shraiman}},\ }\href@noop {} {\bibfield  {journal}
  {\bibinfo  {journal} {Proceedings of the National Academy of Sciences}\
  }\textbf {\bibinfo {volume} {109}},\ \bibinfo {pages} {739} (\bibinfo {year}
  {2012})}\BibitemShut {NoStop}%
\bibitem [{\citenamefont {Eisenhoffer}\ and\ \citenamefont
  {Rosenblatt}(2013)}]{eisenhoffer2013bringing}%
  \BibitemOpen
  \bibfield  {author} {\bibinfo {author} {\bibfnamefont {G.~T.}\ \bibnamefont
  {Eisenhoffer}}\ and\ \bibinfo {author} {\bibfnamefont {J.}~\bibnamefont
  {Rosenblatt}},\ }\href@noop {} {\bibfield  {journal} {\bibinfo  {journal}
  {Trends in cell biology}\ }\textbf {\bibinfo {volume} {23}},\ \bibinfo
  {pages} {185} (\bibinfo {year} {2013})}\BibitemShut {NoStop}%
\bibitem [{\citenamefont {Marinari}\ \emph {et~al.}(2012)\citenamefont
  {Marinari}, \citenamefont {Mehonic}, \citenamefont {Curran}, \citenamefont
  {Gale}, \citenamefont {Duke},\ and\ \citenamefont {Baum}}]{marinari2012live}%
  \BibitemOpen
  \bibfield  {author} {\bibinfo {author} {\bibfnamefont {E.}~\bibnamefont
  {Marinari}}, \bibinfo {author} {\bibfnamefont {A.}~\bibnamefont {Mehonic}},
  \bibinfo {author} {\bibfnamefont {S.}~\bibnamefont {Curran}}, \bibinfo
  {author} {\bibfnamefont {J.}~\bibnamefont {Gale}}, \bibinfo {author}
  {\bibfnamefont {T.}~\bibnamefont {Duke}},\ and\ \bibinfo {author}
  {\bibfnamefont {B.}~\bibnamefont {Baum}},\ }\href@noop {} {\bibfield
  {journal} {\bibinfo  {journal} {Nature}\ }\textbf {\bibinfo {volume} {484}},\
  \bibinfo {pages} {542} (\bibinfo {year} {2012})}\BibitemShut {NoStop}%
\bibitem [{\citenamefont {Schiefelbein}\ \emph {et~al.}(2014)\citenamefont
  {Schiefelbein}, \citenamefont {Huang},\ and\ \citenamefont
  {Zheng}}]{schiefelbein2014regulation}%
  \BibitemOpen
  \bibfield  {author} {\bibinfo {author} {\bibfnamefont {J.}~\bibnamefont
  {Schiefelbein}}, \bibinfo {author} {\bibfnamefont {L.}~\bibnamefont
  {Huang}},\ and\ \bibinfo {author} {\bibfnamefont {X.}~\bibnamefont {Zheng}},\
  }\href@noop {} {\bibfield  {journal} {\bibinfo  {journal} {Frontiers in plant
  science}\ }\textbf {\bibinfo {volume} {5}},\ \bibinfo {pages} {47} (\bibinfo
  {year} {2014})}\BibitemShut {NoStop}%
\bibitem [{\citenamefont {Rauch}\ \emph {et~al.}(2016)\citenamefont {Rauch},
  \citenamefont {Rukhlenko}, \citenamefont {Kolch},\ and\ \citenamefont
  {Kholodenko}}]{rauch2016mapk}%
  \BibitemOpen
  \bibfield  {author} {\bibinfo {author} {\bibfnamefont {N.}~\bibnamefont
  {Rauch}}, \bibinfo {author} {\bibfnamefont {O.~S.}\ \bibnamefont
  {Rukhlenko}}, \bibinfo {author} {\bibfnamefont {W.}~\bibnamefont {Kolch}},\
  and\ \bibinfo {author} {\bibfnamefont {B.~N.}\ \bibnamefont {Kholodenko}},\
  }\href@noop {} {\bibfield  {journal} {\bibinfo  {journal} {Current opinion in
  structural biology}\ }\textbf {\bibinfo {volume} {41}},\ \bibinfo {pages}
  {151} (\bibinfo {year} {2016})}\BibitemShut {NoStop}%
\bibitem [{\citenamefont {Chan}\ \emph {et~al.}(2017)\citenamefont {Chan},
  \citenamefont {Heisenberg},\ and\ \citenamefont
  {Hiiragi}}]{chan2017coordination}%
  \BibitemOpen
  \bibfield  {author} {\bibinfo {author} {\bibfnamefont {C.~J.}\ \bibnamefont
  {Chan}}, \bibinfo {author} {\bibfnamefont {C.-P.}\ \bibnamefont
  {Heisenberg}},\ and\ \bibinfo {author} {\bibfnamefont {T.}~\bibnamefont
  {Hiiragi}},\ }\href@noop {} {\bibfield  {journal} {\bibinfo  {journal}
  {Current Biology}\ }\textbf {\bibinfo {volume} {27}},\ \bibinfo {pages}
  {R1024} (\bibinfo {year} {2017})}\BibitemShut {NoStop}%
\bibitem [{\citenamefont {Greulich}\ and\ \citenamefont
  {Simons}(2016)}]{greulich2016dynamic}%
  \BibitemOpen
  \bibfield  {author} {\bibinfo {author} {\bibfnamefont {P.}~\bibnamefont
  {Greulich}}\ and\ \bibinfo {author} {\bibfnamefont {B.~D.}\ \bibnamefont
  {Simons}},\ }\href@noop {} {\bibfield  {journal} {\bibinfo  {journal}
  {Proceedings of the National Academy of Sciences}\ }\textbf {\bibinfo
  {volume} {113}},\ \bibinfo {pages} {7509} (\bibinfo {year}
  {2016})}\BibitemShut {NoStop}%
\bibitem [{\citenamefont {Kok}\ \emph {et~al.}(2022)\citenamefont {Kok},
  \citenamefont {Tans},\ and\ \citenamefont {van Zon}}]{kok2022minimizing}%
  \BibitemOpen
  \bibfield  {author} {\bibinfo {author} {\bibfnamefont {R.~N.~U.}\
  \bibnamefont {Kok}}, \bibinfo {author} {\bibfnamefont {S.~J.}\ \bibnamefont
  {Tans}},\ and\ \bibinfo {author} {\bibfnamefont {J.~S.}\ \bibnamefont {van
  Zon}},\ }\href@noop {} {\bibfield  {journal} {\bibinfo  {journal} {bioRxiv}\
  } (\bibinfo {year} {2022})}\BibitemShut {NoStop}%
\bibitem [{\citenamefont {Huelsz-Prince}\ \emph {et~al.}(2022)\citenamefont
  {Huelsz-Prince}, \citenamefont {Kok}, \citenamefont {Goos}, \citenamefont
  {Bruens}, \citenamefont {Zheng}, \citenamefont {Ellenbroek}, \citenamefont
  {Van~Rheenen}, \citenamefont {Tans},\ and\ \citenamefont {van
  Zon}}]{huelsz2022mother}%
  \BibitemOpen
  \bibfield  {author} {\bibinfo {author} {\bibfnamefont {G.}~\bibnamefont
  {Huelsz-Prince}}, \bibinfo {author} {\bibfnamefont {R.~N.~U.}\ \bibnamefont
  {Kok}}, \bibinfo {author} {\bibfnamefont {Y.}~\bibnamefont {Goos}}, \bibinfo
  {author} {\bibfnamefont {L.}~\bibnamefont {Bruens}}, \bibinfo {author}
  {\bibfnamefont {X.}~\bibnamefont {Zheng}}, \bibinfo {author} {\bibfnamefont
  {S.}~\bibnamefont {Ellenbroek}}, \bibinfo {author} {\bibfnamefont
  {J.}~\bibnamefont {Van~Rheenen}}, \bibinfo {author} {\bibfnamefont
  {S.}~\bibnamefont {Tans}},\ and\ \bibinfo {author} {\bibfnamefont {J.~S.}\
  \bibnamefont {van Zon}},\ }\href@noop {} {\bibfield  {journal} {\bibinfo
  {journal} {Elife}\ }\textbf {\bibinfo {volume} {11}},\ \bibinfo {pages}
  {e80682} (\bibinfo {year} {2022})}\BibitemShut {NoStop}%
\bibitem [{\citenamefont {Gardiner}\ \emph {et~al.}(1985)\citenamefont
  {Gardiner} \emph {et~al.}}]{gardiner1985handbook}%
  \BibitemOpen
  \bibfield  {author} {\bibinfo {author} {\bibfnamefont {C.~W.}\ \bibnamefont
  {Gardiner}} \emph {et~al.},\ }\href@noop {} {\emph {\bibinfo {title}
  {Handbook of stochastic methods}}},\ Vol.~\bibinfo {volume} {3}\ (\bibinfo
  {publisher} {springer Berlin},\ \bibinfo {year} {1985})\BibitemShut {NoStop}%
\bibitem [{\citenamefont {Stein}(2012)}]{stein2012transport}%
  \BibitemOpen
  \bibfield  {author} {\bibinfo {author} {\bibfnamefont {W.}~\bibnamefont
  {Stein}},\ }\href@noop {} {\emph {\bibinfo {title} {Transport and diffusion
  across cell membranes}}}\ (\bibinfo  {publisher} {Elsevier},\ \bibinfo {year}
  {2012})\BibitemShut {NoStop}%
\bibitem [{\citenamefont {Dal~Co}\ \emph {et~al.}(2017)\citenamefont {Dal~Co},
  \citenamefont {Lagomarsino}, \citenamefont {Caselle},\ and\ \citenamefont
  {Osella}}]{co2017stochastic}%
  \BibitemOpen
  \bibfield  {author} {\bibinfo {author} {\bibfnamefont {A.}~\bibnamefont
  {Dal~Co}}, \bibinfo {author} {\bibfnamefont {M.~C.}\ \bibnamefont
  {Lagomarsino}}, \bibinfo {author} {\bibfnamefont {M.}~\bibnamefont
  {Caselle}},\ and\ \bibinfo {author} {\bibfnamefont {M.}~\bibnamefont
  {Osella}},\ }\href@noop {} {\bibfield  {journal} {\bibinfo  {journal}
  {Nucleic acids research}\ }\textbf {\bibinfo {volume} {45}},\ \bibinfo
  {pages} {1069} (\bibinfo {year} {2017})}\BibitemShut {NoStop}%
\bibitem [{\citenamefont {Alon}(2006)}]{alon2006introduction}%
  \BibitemOpen
  \bibfield  {author} {\bibinfo {author} {\bibfnamefont {U.}~\bibnamefont
  {Alon}},\ }\href@noop {} {\emph {\bibinfo {title} {An introduction to systems
  biology: design principles of biological circuits}}}\ (\bibinfo  {publisher}
  {Chapman and Hall/CRC},\ \bibinfo {year} {2006})\BibitemShut {NoStop}%
\bibitem [{\citenamefont {Gillespie}(1976)}]{gillespie1976general}%
  \BibitemOpen
  \bibfield  {author} {\bibinfo {author} {\bibfnamefont {D.~T.}\ \bibnamefont
  {Gillespie}},\ }\href@noop {} {\bibfield  {journal} {\bibinfo  {journal}
  {Journal of computational physics}\ }\textbf {\bibinfo {volume} {22}},\
  \bibinfo {pages} {403} (\bibinfo {year} {1976})}\BibitemShut {NoStop}%
\bibitem [{\citenamefont {Erban}\ \emph {et~al.}(2007)\citenamefont {Erban},
  \citenamefont {Chapman},\ and\ \citenamefont {Maini}}]{erban2007practical}%
  \BibitemOpen
  \bibfield  {author} {\bibinfo {author} {\bibfnamefont {R.}~\bibnamefont
  {Erban}}, \bibinfo {author} {\bibfnamefont {J.}~\bibnamefont {Chapman}},\
  and\ \bibinfo {author} {\bibfnamefont {P.}~\bibnamefont {Maini}},\
  }\href@noop {} {\bibfield  {journal} {\bibinfo  {journal} {arXiv preprint
  arXiv:0704.1908}\ } (\bibinfo {year} {2007})}\BibitemShut {NoStop}%
\bibitem [{\citenamefont {Gagliardi}\ \emph {et~al.}(2021)\citenamefont
  {Gagliardi}, \citenamefont {Dobrzy{\'n}ski}, \citenamefont {Jacques},
  \citenamefont {Dessauges}, \citenamefont {Ender}, \citenamefont {Blum},
  \citenamefont {Hughes}, \citenamefont {Cohen},\ and\ \citenamefont
  {Pertz}}]{gagliardi2021collective}%
  \BibitemOpen
  \bibfield  {author} {\bibinfo {author} {\bibfnamefont {P.~A.}\ \bibnamefont
  {Gagliardi}}, \bibinfo {author} {\bibfnamefont {M.}~\bibnamefont
  {Dobrzy{\'n}ski}}, \bibinfo {author} {\bibfnamefont {M.-A.}\ \bibnamefont
  {Jacques}}, \bibinfo {author} {\bibfnamefont {C.}~\bibnamefont {Dessauges}},
  \bibinfo {author} {\bibfnamefont {P.}~\bibnamefont {Ender}}, \bibinfo
  {author} {\bibfnamefont {Y.}~\bibnamefont {Blum}}, \bibinfo {author}
  {\bibfnamefont {R.~M.}\ \bibnamefont {Hughes}}, \bibinfo {author}
  {\bibfnamefont {A.~R.}\ \bibnamefont {Cohen}},\ and\ \bibinfo {author}
  {\bibfnamefont {O.}~\bibnamefont {Pertz}},\ }\href@noop {} {\bibfield
  {journal} {\bibinfo  {journal} {Developmental cell}\ }\textbf {\bibinfo
  {volume} {56}},\ \bibinfo {pages} {1712} (\bibinfo {year}
  {2021})}\BibitemShut {NoStop}%
\bibitem [{\citenamefont {Valon}\ \emph {et~al.}(2021)\citenamefont {Valon},
  \citenamefont {Davidovi{\'c}}, \citenamefont {Levillayer}, \citenamefont
  {Villars}, \citenamefont {Chouly}, \citenamefont {Cerqueira-Campos},\ and\
  \citenamefont {Levayer}}]{valon2021robustness}%
  \BibitemOpen
  \bibfield  {author} {\bibinfo {author} {\bibfnamefont {L.}~\bibnamefont
  {Valon}}, \bibinfo {author} {\bibfnamefont {A.}~\bibnamefont
  {Davidovi{\'c}}}, \bibinfo {author} {\bibfnamefont {F.}~\bibnamefont
  {Levillayer}}, \bibinfo {author} {\bibfnamefont {A.}~\bibnamefont {Villars}},
  \bibinfo {author} {\bibfnamefont {M.}~\bibnamefont {Chouly}}, \bibinfo
  {author} {\bibfnamefont {F.}~\bibnamefont {Cerqueira-Campos}},\ and\ \bibinfo
  {author} {\bibfnamefont {R.}~\bibnamefont {Levayer}},\ }\href@noop {}
  {\bibfield  {journal} {\bibinfo  {journal} {Developmental Cell}\ }\textbf
  {\bibinfo {volume} {56}},\ \bibinfo {pages} {1700} (\bibinfo {year}
  {2021})}\BibitemShut {NoStop}%
\bibitem [{\citenamefont {Mesa}\ \emph {et~al.}(2018)\citenamefont {Mesa},
  \citenamefont {Kawaguchi}, \citenamefont {Cockburn}, \citenamefont
  {Gonzalez}, \citenamefont {Boucher}, \citenamefont {Xin}, \citenamefont
  {Klein},\ and\ \citenamefont {Greco}}]{mesa2018homeostatic}%
  \BibitemOpen
  \bibfield  {author} {\bibinfo {author} {\bibfnamefont {K.~R.}\ \bibnamefont
  {Mesa}}, \bibinfo {author} {\bibfnamefont {K.}~\bibnamefont {Kawaguchi}},
  \bibinfo {author} {\bibfnamefont {K.}~\bibnamefont {Cockburn}}, \bibinfo
  {author} {\bibfnamefont {D.}~\bibnamefont {Gonzalez}}, \bibinfo {author}
  {\bibfnamefont {J.}~\bibnamefont {Boucher}}, \bibinfo {author} {\bibfnamefont
  {T.}~\bibnamefont {Xin}}, \bibinfo {author} {\bibfnamefont {A.~M.}\
  \bibnamefont {Klein}},\ and\ \bibinfo {author} {\bibfnamefont
  {V.}~\bibnamefont {Greco}},\ }\href@noop {} {\bibfield  {journal} {\bibinfo
  {journal} {Cell stem cell}\ }\textbf {\bibinfo {volume} {23}},\ \bibinfo
  {pages} {677} (\bibinfo {year} {2018})}\BibitemShut {NoStop}%
\bibitem [{\citenamefont {Rompolas}\ \emph {et~al.}(2016)\citenamefont
  {Rompolas}, \citenamefont {Mesa}, \citenamefont {Kawaguchi}, \citenamefont
  {Park}, \citenamefont {Gonzalez}, \citenamefont {Brown}, \citenamefont
  {Boucher}, \citenamefont {Klein},\ and\ \citenamefont
  {Greco}}]{rompolas2016spatiotemporal}%
  \BibitemOpen
  \bibfield  {author} {\bibinfo {author} {\bibfnamefont {P.}~\bibnamefont
  {Rompolas}}, \bibinfo {author} {\bibfnamefont {K.~R.}\ \bibnamefont {Mesa}},
  \bibinfo {author} {\bibfnamefont {K.}~\bibnamefont {Kawaguchi}}, \bibinfo
  {author} {\bibfnamefont {S.}~\bibnamefont {Park}}, \bibinfo {author}
  {\bibfnamefont {D.}~\bibnamefont {Gonzalez}}, \bibinfo {author}
  {\bibfnamefont {S.}~\bibnamefont {Brown}}, \bibinfo {author} {\bibfnamefont
  {J.}~\bibnamefont {Boucher}}, \bibinfo {author} {\bibfnamefont {A.~M.}\
  \bibnamefont {Klein}},\ and\ \bibinfo {author} {\bibfnamefont
  {V.}~\bibnamefont {Greco}},\ }\href@noop {} {\bibfield  {journal} {\bibinfo
  {journal} {Science}\ }\textbf {\bibinfo {volume} {352}},\ \bibinfo {pages}
  {1471} (\bibinfo {year} {2016})}\BibitemShut {NoStop}%
\bibitem [{\citenamefont {Hart}\ \emph {et~al.}(2012)\citenamefont {Hart},
  \citenamefont {Antebi}, \citenamefont {Mayo}, \citenamefont {Friedman},\ and\
  \citenamefont {Alon}}]{hart2012design}%
  \BibitemOpen
  \bibfield  {author} {\bibinfo {author} {\bibfnamefont {Y.}~\bibnamefont
  {Hart}}, \bibinfo {author} {\bibfnamefont {Y.~E.}\ \bibnamefont {Antebi}},
  \bibinfo {author} {\bibfnamefont {A.~E.}\ \bibnamefont {Mayo}}, \bibinfo
  {author} {\bibfnamefont {N.}~\bibnamefont {Friedman}},\ and\ \bibinfo
  {author} {\bibfnamefont {U.}~\bibnamefont {Alon}},\ }\href@noop {} {\bibfield
   {journal} {\bibinfo  {journal} {Proceedings of the National Academy of
  Sciences}\ }\textbf {\bibinfo {volume} {109}},\ \bibinfo {pages} {8346}
  (\bibinfo {year} {2012})}\BibitemShut {NoStop}%
\bibitem [{\citenamefont {Hart}\ \emph {et~al.}(2014)\citenamefont {Hart},
  \citenamefont {Reich-Zeliger}, \citenamefont {Antebi}, \citenamefont
  {Zaretsky}, \citenamefont {Mayo}, \citenamefont {Alon},\ and\ \citenamefont
  {Friedman}}]{hart2014paradoxical}%
  \BibitemOpen
  \bibfield  {author} {\bibinfo {author} {\bibfnamefont {Y.}~\bibnamefont
  {Hart}}, \bibinfo {author} {\bibfnamefont {S.}~\bibnamefont {Reich-Zeliger}},
  \bibinfo {author} {\bibfnamefont {Y.~E.}\ \bibnamefont {Antebi}}, \bibinfo
  {author} {\bibfnamefont {I.}~\bibnamefont {Zaretsky}}, \bibinfo {author}
  {\bibfnamefont {A.~E.}\ \bibnamefont {Mayo}}, \bibinfo {author}
  {\bibfnamefont {U.}~\bibnamefont {Alon}},\ and\ \bibinfo {author}
  {\bibfnamefont {N.}~\bibnamefont {Friedman}},\ }\href@noop {} {\bibfield
  {journal} {\bibinfo  {journal} {Cell}\ }\textbf {\bibinfo {volume} {158}},\
  \bibinfo {pages} {1022} (\bibinfo {year} {2014})}\BibitemShut {NoStop}%
\end{thebibliography}%

\end{document}